\documentclass{aa}  
\pdfoutput=1
\usepackage{graphicx}
\usepackage{txfonts}
\usepackage{color}
\definecolor{blue}{RGB}{0,0,255}
\usepackage{tabu}
\usepackage{tikz}
\usepackage{multirow}

\begin{document}

   \title{From Stellar Coron\ae\ to Gyrochronology: a theoretical and observational exploration}

   \subtitle{}

   \author{J. Ahuir, A. S. Brun \and A. Strugarek}

   \institute{D\'epartement d'Astrophysique-AIM, CEA/DRF/IRFU, CNRS/INSU, Universit\'e Paris-Saclay, Universit\'e Paris-Diderot, Universit\'e de Paris, F-91191 Gif-sur-Yvette, France\\
              \email{jeremy.ahuir@cea.fr}
             }

   \date{Received XXX ; accepted YYY}

 
  \abstract
   {Stellar spin-down is the result of a complex process involving rotation, dynamo, wind and magnetism. Multi-wavelength surveys of solar-like stars have revealed the likely existence of relationships between their rotation, X-ray luminosity, mass-losses and magnetism. Those impose strong constraints on the corona and wind of cool stars.}
   {We aim to provide power-law prescriptions of the mass-loss of stars, of their magnetic field, and of their base coronal density and temperature that are compatible with their observationally-constrained spin-down.}
   {We link the magnetic field and the mass-loss rate from a wind torque formulation in agreement with the distribution of stellar rotation periods in open clusters and the Skumanich law. Given a wind model and an expression of the X-ray luminosity from radiative losses, we constrain the coronal properties by assuming different physical scenarii linking closed loops to coronal holes.}
   {We find that the magnetic field and the mass loss are involved in a one-to-one correspondence constrained from spin-down considerations. We show that a magnetic field depending on both the Rossby number and the stellar mass is required to keep a consistent spin-down model. The estimates of the magnetic field and the mass-loss rate obtained from our formalism are consistent with statistical studies as well as individual observations and give new leads to constrain the magnetic field-rotation relation. The set of scaling-laws we derived can be broadly applied to cool stars from the PMS to the end of the MS, and allow for a stellar wind modelling consistent with all the observational constraints available to date.}
   {}


   \maketitle
%

\section{Introduction}

The rotation of stars is subject to a complex evolution along their life. During the early stage of their lifetime, solar-type stars spin-up as they contract during the Pre-Main Sequence. Once the Zero-Age Main Sequence (ZAMS) is reached, they keep their moment of inertia relatively constant \citep{armitage} while they lose mass and angular momentum through the flow of a magnetized stellar wind \citep{schatzman,webdav,mestel68}. It results in a slow-down of their rotation as they age, which approximately follows the empirical Skumanich’s law: \(\Omega_\star \propto t^{-0.5}\) \citep{skumanich72}. This spin-down also depends on the stellar mass \citep{webdav,matt15}. Indeed, during most of the Pre-Main Sequence, lower mass stars tend to remain fast rotators for a longer time than the higher mass stars. Then, after hundreds of millions of years, the slowest rotators converge toward a sequence in which the rotation rate increases with mass. These phenomena make gyrochronology possible \citep{barnes03}, thereby allowing the estimation of stellar ages through measurements of rotation periods and masses.

Understanding the feedback loop between rotation, dynamo action, magnetism and wind is a key issue to predict the behavior of the solar-like stars as they evolve. It is also important to understand the evolution of star-planet systems and to follow potential planetary migrations \citep{zhangpenev,benbakoura}. Our ability to track the stellar rotation evolution strongly relies on the wind braking modeling. Most angular momentum evolution models fall back on the \citet{kawaler} prescription to assess the wind torque, which is expressed in this model as a power-law depending on the magnetic field, the mass-loss rate, the mass and the radius of the star. Several modifications have since been brought to this formulation \citep{krishnamurthi,bouvier97,reinersmohanty}. For instance, to account for fast rotators on the ZAMS, a saturation of the braking torque is required \citep{barnessofia96}. Magnetohydrodynamical simulations can also be used to assess of the angular momentum loss due to the wind. For instance, \citet{matt12}, following \citet{mattpudritz}, simulated the flow of a stellar wind along the opened field lines of a dipolar configuration to estimate the torque, by taking into account the influence of stellar rotation on the wind acceleration \citep{sakurai}. More recently, modified versions of this formulation were presented to take into account the influence of the magnetic topology. \citet{reville15a} considered the magnetic flux through the open field lines to build a topology-independant wind torque, while \citet{finleymatt17,finleymatt18} relied on a broken power-law behavior to deal with combined geometries. \citet{garraffo16,garraffo18} accounted for magnetic topology by means of a modulating factor to the angular momentum loss estimated for a dipolar configuration. In general, in most of the prescriptions of spin-evolution torques, it is assumed that the wind carries away angular momentum at a rate proportional to \(\Omega_\star^3\) during the Main Sequence in order to follow the Skumanich law. It is important to note that the Skumanich law is today questioned for evolved stars. Some recent studies have shown a substantial decrease of the wind braking efficiency for evolved stars around the solar value of the Rossby number \citep{vansaders}, even if this alternative scenario seems to be in disagreement with solar twins studies \citep{lorenzo-oliveira}. 

To estimate the angular momentum loss, we need to know accurately the stellar magnetic field and the properties of its wind (such as the mass loss induced). The latter are bound to physical parameters in the corona, like the plasma temperature and density. Several observational trends constrain and correlate those different quantities: the mass loss \citep{wood02,wood05,jardine19}, the X-ray activity \citep{pizzolato,wright,reiners}, the magnetic field \citep{vidotto14,see17} and the rotation rate of the star for different ages \citep{agueros,mcquillan14,galletbouvier15}. Coupling all those quantities is therefore necessary to design a consistent model of stellar spin-down. 

A preliminary exploration has been carried out by \citet{blackman}, who presented a simplified model for the coupled time evolution of the relevant quantities on the basis of a pressure-driven isothermal wind \citep{parker}. In this framework, a dynamo-induced magnetic field dictates the behavior of the stellar wind and the X-ray luminosity through a coronal equilibrium. More recently, \citet{skumanich19} focused on the connexions between the physical parameters regulating the stellar spin-down by considering their rotational evolution. By assuming a direct correlation between the mass loss and the magnetic field, he studied the influence of the Skumanich law on all the relevant quantities by means on a variety of observational trends. Such a study suggest that the magnetic field and the rotation of cool star should scale linearly, while the mass-loss should scale quadratically with stellar rotation.

More generally, stellar wind models require the knowledge of the coronal temperature and density. To be consistent with the aforementioned correlations, one need to constrain those two quantities from tracers of the coronal activity, like the soft X-ray emission of the star. In this spirit, by relying on a 1D polytropic and magnetocentrifugal wind, \citet{holzwarth} provided scaling laws in accordance with the rotational evolution of the X-ray luminosity \citep{ivanova} and the empirical mass loss-X-ray flux correlation from \citet{wood05}.\\

Building on those previous studies, the main goal of this paper is to infer from stellar spin-down considerations some prescriptions of the magnetic field, the mass loss, the coronal temperature and the coronal density as a function of fundamental stellar parameters (such as mass, radius and rotation rate) in order to be consistent with all the observational trends. Each of those constraints are successively introduced to eliminate the largest number of free parameters involved in the spin-down process. Please note that in our attempt to extract the most important interdependencies between various physical mechanisms involved in stellar spin down theory, we had to make some simplifying assumptions but we have been careful to retain all the key mechanisms. Our scaling laws provide a novel and systematic way to connect all these mechanisms together and should be seen as a first approach to systematically characterize these complex relationships between the stellar parameters. In Section 2 we introduce the theoretical framework we used for the torque parametrization. In Section 3 observational constraints are leveraged to unveil the inter-dependency of the magnetic field and the mass-loss rate of cool stars. We further derive the associated prescriptions for the coronal temperature and the coronal density for a given wind model. In Section 4, we summarize our prescriptions and give a practical application to the case of \(\epsilon\) Eridani. All those results are then summarized, discussed and put in perspective in Section 5.


\section{Stellar wind torques of solar-type stars: theoretical approach}
\subsection{Fundamental stellar parameters and architecture of the model}

First of all, designing a consistent model for the stellar spin-down requires to inventory the coupling between the various physical mechanisms (and their associated control parameters) involved in the process.

\begin{figure*}[!h]
\centering
\includegraphics[scale=0.44]{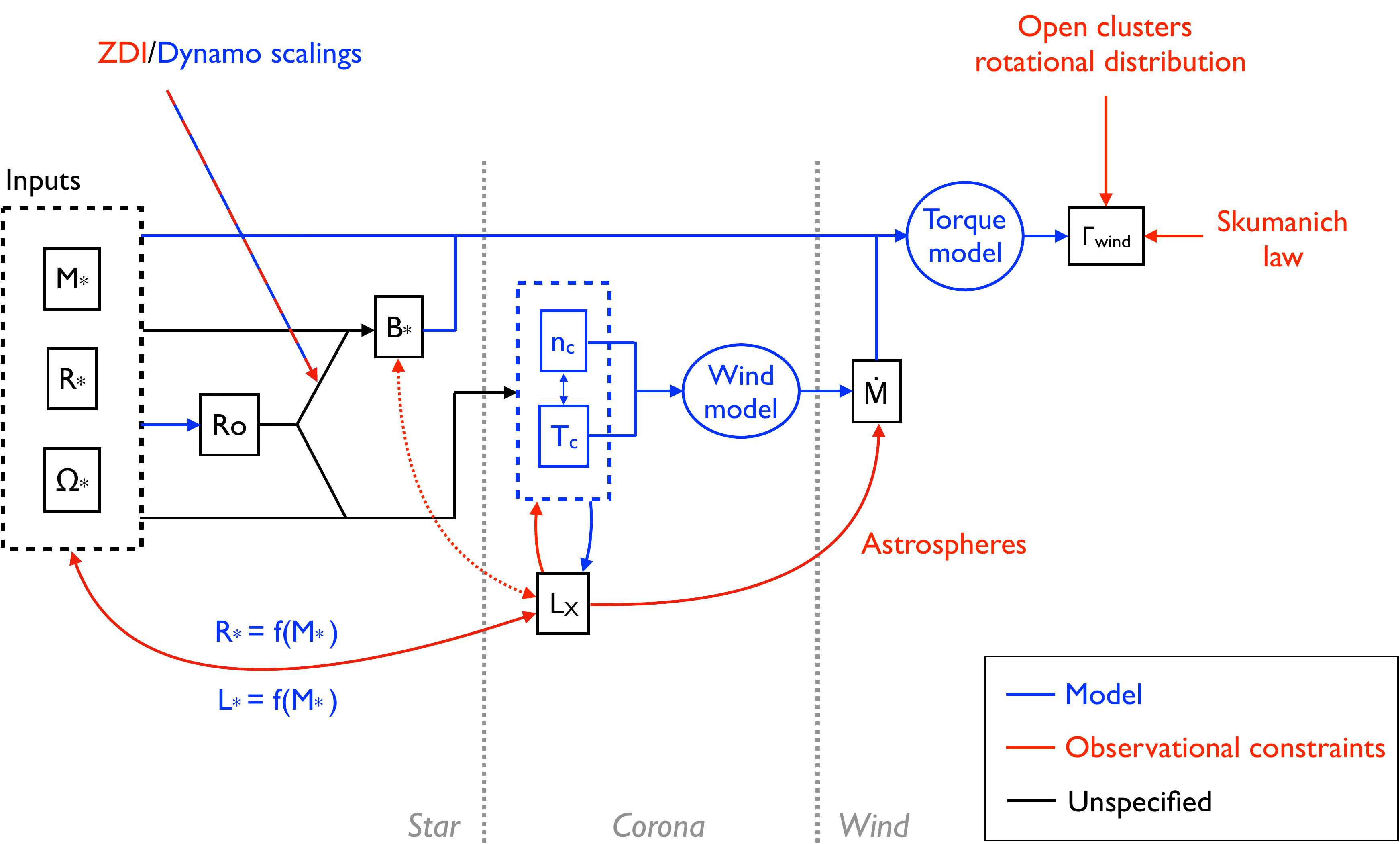}
\caption{Sketch of the multiple couplings between the inputs of the model (stellar mass, radius and rotation rate), the magnetic field, the X-ray luminosity, the coronal temperature and density, the mass loss and the wind braking torque. In blue: model assumptions. In red: observational constraints.}
\label{interdep}
\end{figure*}

Stellar rotation and magnetism strengthen the supersonic flow of a stellar wind \citep{webdav}, which entails itself a mass loss \(\dot M\). The wind then carries away angular momentum, leading to a braking torque \(\Gamma_\text{wind}\) spinning down the star. The logical sequence from fundamental stellar parameters to the wind braking torque and the role of the X-ray luminosity inside the architecture of the model is summarized in Figure \ref{interdep}, where we show all the interdependencies between the various physical mechanisms and their control parameters. Theoretical assumptions are represented in blue and are essentially related to the wind torque parametrization, dynamo scalings as well as the choice of a wind model. Observational trends, in red, are taken as constraints for the physical models (see Appendix F for an extensive view of those ingredients and their caveats). 

Solar-type stars generate a magnetic field through a dynamo in their envelope \citep{brun04,brun15} triggered by turbulent convective movements. Those can be influenced by stellar rotation \citep{durney}. This effect can be quantified by the Rossby number \(Ro\), which has been shown to be a relevant quantity to characterize the magnetic activity of the stars \citep{noyes}. Direct spectropolarimetric studies also exhibit scaling laws linking the stellar magnetic field and this dimensionless number \citep{vidotto14,see17}. However, multiple definitions of the Rossby number are available in the literature, which has has forced the community to clarify this aspect \citep{landin,brun17,amard19}. In this work, the Rossby number will be normalized to the solar value. This way, different prescriptions of the Rossby number can be equally used as long as they differ only by a proportionality factor. This is generally the case of main-sequence stars, which we will focus on here. However, to deal with individual stars, we will use for simplicity the stellar Rossby number, expressed as
\begin{equation}\label{eqn:def_rossby}
Ro = \frac{P_\text{rot}}{\tau_c},
\end{equation}
with \(P_\text{rot}\) the stellar rotation period and \(\tau_{c}\) the convective turnover time. The latter can itself be assessed at different locations or computed in a variety of ways, which are often linked to a given stellar evolution model \citep{landin,cranmer,ardestani} or a set of observations \citep{wright}.

Coronal activity, measured with the X-ray emission of the star, has been correlated with the stellar magnetic field \citep{petsov,vidotto14} and the mass loss \citep{wood02,wood05}. Observations have also exhibited a relationship between the stellar rotation and the X-ray luminosity for main sequence stars \citep{noyes,pizzolato,wright,vidotto14,reiners}, allowing us to link the coronal activity to fundamental stellar parameters. This gives us the opportunity to closely link those physical quantities together through stellar activity considerations.

Furthermore, the corona of the stars we consider is thought to be heated up through magnetic processes  \citep[for a review, see][]{mathioudakis,cranmer17} to reach a typical coronal temperature \(T_c\) of around \(10^6\) K \citep{suzuki06}. Such a hot and diluted medium (with a density \(n_c\)) expands and generates a transsonic wind \citep{parker}. Observationally, scaling laws have been established between the coronal temperature and the X-ray emission of the star \citep{preibisch97,johnstonegudel,wood18}. From a theoretical point of view, the latter can be deduced from coronal properties by estimating the radiative losses in the X-ray–emitting area of the corona \citep{ivanova,aschwanden,blackman}. Therefore, it could be possible to provide a prescription for the coronal temperature and density in terms of stellar mass, radius and rotation rate.

\subsection{Torque parametrization}
Solar-like stars spin down due to the angular momentum extraction by the stellar wind \citep{schatzman}.
\citet{webdav} showed with a 1D model in the stellar equatorial plane that the angular momentum loss \(\Gamma_\text{wind}\) can be estimated at the Alfv\'en radius \(r_A\) as
\begin{equation}
\Gamma_{wind} = \dot{M}r_A^2\Omega_\star,    
\end{equation}
with \(\Omega_\star\) the stellar rotation rate and \(\dot M\) the mass loss.
The Alfv\'en radius can be expressed as a function of the stellar magnetic field, the mass loss and other fundamental stellar parameters, such as the stellar mass, the stellar radius and the stellar rotation rate \citep{kawaler}. For instance, \citet{matt12} presented the following expression for this characteristic distance, by assuming a dipolar magnetic field:
\begin{equation}
\frac{\left<r_A\right>}{R_\star}=K_1\left[\frac{\Upsilon}{\sqrt{K_2^2+0.5f^2}}\right]^m,
\end{equation}
where \(M_\star\) is the stellar mass, \(R_\star\) the stellar radius, \(K_1, K_2\ \text{and}\ m\) are constants set using 2D MHD simulations. More precisely, \(K_1\) is used to calibrate the solar wind torque, \(K_2\) the efficiency of the magnetocentrifugal acceleration \citep{sakurai}, and \(m\) stands for a magnetic topology parameter \citep{reville15a}. For a dipolar field, we will use the value proposed in  \citet{matt12}, \textit{i.e.} \(m=0.2177\). \(f = \Omega_\star/\sqrt{GM_\star/R_\star^3}\) is the break-up ratio, obtained by dividing the stellar rotation rate at the equator of the star by the keplerian angular velocity. The magnetization parameter \(\Upsilon\) \citep{mattpudritz} is defined as
\begin{equation}
\Upsilon = \frac{B_\star^2R_\star^2}{\dot{M}v_\text{esc}},
\end{equation}
with \(B_\star\) the magnetic field strength at the stellar equator and \(v_\text{esc}=\sqrt{2GM_\star/R_\star}\) the escape velocity. With this formulation, the wind braking torque becomes
\begin{equation}
\begin{split}
\Gamma_\text{wind} &= \dot{M}\Omega_\star R_\star^2 K_1^2\left[\frac{\Upsilon}{\sqrt{K_2^2+0.5f^2}}\right]^{2m}\\
&\propto \dot{M}^{1-2m}B_\star^{4m} R_\star^{2+5m}M_\star^{-m}\Omega_\star\left(K_2^2+0.5f^2\right)^{-m}.
\end{split}
\end{equation}
We will rewrite this power-law expansion of the torque to adopt the following generic expression
\begin{equation}\label{eqn:torque}
\Gamma_\text{wind} \propto \dot{M}^{1-2m}B_\star^{4m}R_\star^{2+5m}M_\star^{-m}\Omega_\star\left[1+\frac{f^2}{K^2}\right]^{-m},
\end{equation}
where \(K = \sqrt{2}K_2\) is a constant.

\subsection{Stellar magnetic field prescription}
Computing the wind braking torque requires to estimate the stellar magnetic field and the mass-loss rate. We focus here on the surface magnetic field strength at the stellar equator \(B_\star\). For the sake of simplicity we will assume a power-law expression for the magnetic field as a function of the stellar mass, the stellar radius and the Rossby number as 
\begin{equation}\label{eqn:magfield}
B_\star\propto \left(\frac{Ro}{Ro_\odot}\right)^{-p_B}\left(\frac{R_\star}{R_\odot}\right)^{r_B}\left(\frac{M_\star}{M_\odot}\right)^{m_B},
\end{equation}
where \(p_B, m_B\ \&\ r_B\) are unspecified exponents for the time being.

Since we will present in the remaining of this work a significant number of power-law prescriptions, we have to fix a generic notation for the different exponents. More precisely, for the physical quantities we want to estimate (namely the magnetic field, the mass loss and the coronal properties), we will write a given exponent with a lowercase letter indicating the variable of the power law (\(p\) for the rotation period, \(r\) for the stellar radius and \(m\) for the stellar mass). This letter will have a subscript in capital letters representing the quantity for which we give a prescription. As an example, to express a quantity \(A\) as a function of the stellar mass \(M_\star\), we will write
\begin{equation}
A \propto M_\star^{m_A}.
\end{equation}

\subsection{Mass loss prescription}
Because of the wind torque parametrization and the stellar magnetic field prescription, for the sake of consistency we will consider a power-law expression of the stellar mass loss
\begin{equation}\label{eqn:mdot}
\dot M\propto \left(\frac{Ro}{Ro_\odot}\right)^{-p_{\dot M}}\left(\frac{R_\star}{R_\odot}\right)^{r_{\dot M}}\left(\frac{M_\star}{M_\odot}\right)^{m_{\dot M}},
\end{equation}
where \(p_{\dot M}, m_{\dot M}\ \&\ r_{\dot M}\) will be constrained later. Note that those exponents follow the nomenclature presented in Section 2.3. However, the mass loss can be obtained from coronal quantities through a wind model, for example by assuming a radial polytropic pressure-driven outflow, with an index \(\gamma\) (cf. Appendix A):
\begin{equation}\label{eqn:mdotwindfin}
\dot M \propto \left(\frac{M_\star}{M_\odot}\right)^2 \left(\frac{n_c}{n_\odot}\right) \left(\frac{T_c}{T_\odot}\right)^{-\frac{3}{2}}\left[1-\frac{T_{\text{min},\odot}}{T_\odot}\frac{M_\star}{M_\odot}\frac{R_\odot}{R_\star}\frac{T_\odot}{T_c}\right]^{\frac{5-3\gamma}{2(\gamma-1)}},
\end{equation}
where \(n_\odot\), \(T_\odot\)  are respectively the solar values of the density and temperature at the base of the corona. \(T_{\text{min},\odot}=(1-1/\gamma)\ Gm_p M_\odot/2k_B R_\odot\approx 11\  (1-1/\gamma)\) MK  is the minimal temperature needed at the base of the corona
for the Sun to obtain a transsonic wind. For instance, for \(\gamma = 1.05\), value commonly used in the literature
\citep{matt12,reville15a,finleymatt17}, we have \( T_{\text{min},\odot} \approx 0.52\ \text{MK}\). Such an expression will be used in Section 3.5.

\subsection{General formulation of the torque}
The parametrization of the torque with equation \eqref{eqn:torque}, together with the power-law expressions of the stellar magnetic field and the mass loss (equations \eqref{eqn:magfield} and \eqref{eqn:mdot}) lead to the following formulation of the torque as a function of fundamental stellar parameters
\begin{equation}\label{eqn:master}
\begin{split}
\Gamma_\text{wind} \propto \ Ro^{-4m.p_B-(1-2m)p_{\dot M}}&R_\star^{2+5m+4m.r_B+(1-2m)r_{\dot M}}\ \times\\
&M_\star^{-m+4m.m_B+(1-2m)m_{\dot M}}\frac{\Omega_\star}{\left(1+\frac{f^2}{K^2}\right)^m}.
\end{split}
\end{equation}
From this generic formulation, it is now possible to constrain the different exponents by taking into account several observational trends.

\section{Observational constraints}
\subsection{Relationships between stellar parameters}

Some observational trends are based on a set of main-sequence stars in the unsaturated rotation regime (for which the Rossby number is greater than a certain threshold) and therefore take into account scaling laws between stellar parameters, depending on a specific stellar evolution model. In particular we need to consider mass-radius and mass-luminosity relationships during the Main Sequence. We assume here the general correlations
\begin{equation}\label{eqn:LM}
L_\star\propto M_\star^\eta
\end{equation}
\begin{equation}\label{eqn:RM}
R_\star \propto M_\star^\xi,
\end{equation}
where \(\eta\ \text{and}\ \xi\) are constants which depend on which stellar model is considered. Therefore, the upcoming formulations, only valid during the Main Sequence, can accommodate any evolutionary model. We will by default use the typical values \(\eta = 4\) and \(\xi=0.9\) \citep{kippenhahn}.

\subsection{Constraints form stellar rotational evolution}
As already introduced, stellar spin-down studies in open clusters (first performed on the Pleiades, Ursa Major, and the Hyades) show that the rotation rate of evolved main sequence stars tends to converge to the solar rate on a sequence where \(\Omega_\star \propto t^{-1/2}\) \citep{skumanich72}. By assuming that the moment of inertia of the star is constant, the wind braking torque is constrained in the unsaturated regime to scale as the cube of the stellar rotation rate
\begin{equation}\label{eqn:torquesku}
\Gamma_\text{wind} \propto \Omega_\star^3.
\end{equation}
In the following sections, we will assume that gyrochonology is  valid thanks to the Skumanich law. This way, if a decrease of the wind braking efficiency is genuinely happening for evolved stars \citep{vansaders}, then we will only consider solar-type stars younger than the Sun to ensure the validity of equation \eqref{eqn:torquesku}. Note that a stalling of the magnetic braking could be modelled in our formalism with a \textit{re-saturation} regime where the Rossby number is greater than a certain threshold \(Ro_\text{break}\).   

\citet{matt15} studied in more details the different dependencies of the wind braking torque to explain some characteristic features of the distribution of stellar rotation periods in open clusters and \textit{Kepler} stars as a function of their mass. To this end, they focus on two kinds of stellar populations: the slow rotators, in an unsaturated regime, and the fast saturated rotators. The saturation threshold is given by a value of the Rossby number \(Ro_\text{sat}\), which is assumed to be independent of any stellar parameters, at least at zeroth-order. To explain the mass dependency of the stellar spin-down, they take into account the Skumanich law in the unsaturated regime and a linear saturation for the wind braking torque, leading to the following prescription
\begin{equation}\label{eqn:matt15unsat}
\Gamma_\text{wind} = \Gamma_\odot\left(\frac{R_\star}{R_\odot}\right)^a\left(\frac{M_\star}{M_\odot}\right)^b \left(\frac{Ro}{Ro_\odot}\right)^{-2}\left(\frac{\Omega_\star}{\Omega_\odot}\right)\quad\text{(unsaturated)}
\end{equation}
\begin{equation}
\Gamma_\text{wind} = \Gamma_\odot\left(\frac{R_\star}{R_\odot}\right)^a\left(\frac{M_\star}{M_\odot}\right)^b \left(\frac{Ro_{sat}}{Ro_\odot}\right)^{-2}\left(\frac{\Omega_\star}{\Omega_\odot}\right) \quad\text{(saturated),}
\end{equation}
with \(\Gamma_\odot = 6.3\times10^{30}\ \text{erg}, a=3.1\ \text{and}\ b=0.5\).

This prescription can be compared with our formulation, in equation \eqref{eqn:master}, by assuming \(f\ll 1\) and neglecting secular changes of the stellar parameters. It is thus possible to link the magnetic field of the star and the mass loss through the following conditions\\
\begin{equation}\label{eqn:cons_rot1}
p_B = \frac{1}{2m}-\frac{1-2m}{4m}p_{\dot M}\qquad\text{(unsaturated)}
\end{equation}
\begin{equation}\label{eqn:cons_rot2}
p_B = -\frac{1-2m}{4m}p_{\dot M}\qquad\text{(saturated)}
\end{equation}
\begin{equation}\label{eqn:cons_rot3}
r_B = \frac{a-(2+5m)}{4m}-\frac{1-2m}{4m}r_{\dot M}
\end{equation}
\begin{equation}\label{eqn:cons_rot4}
m_B = \frac{b+m}{4m}-\frac{1-2m}{4m}m_{\dot M}.
\end{equation}
\\
Equation \eqref{eqn:cons_rot1} is similar to the condition obtained by \citet{skumanich19}, if we take \(p_B = 1/\beta\ \&\ p_{\dot M} = \alpha/\beta\), according to his notation. However, no correlation between \(B_\star\) and \(\dot M\) is assumed here. The wind torque parametrization therefore gives us the opportunity to infer the mass loss prescription from the magnetic field prescription and reciprocally.

\begin{figure}[!h]
\centering
\includegraphics[scale=0.325]{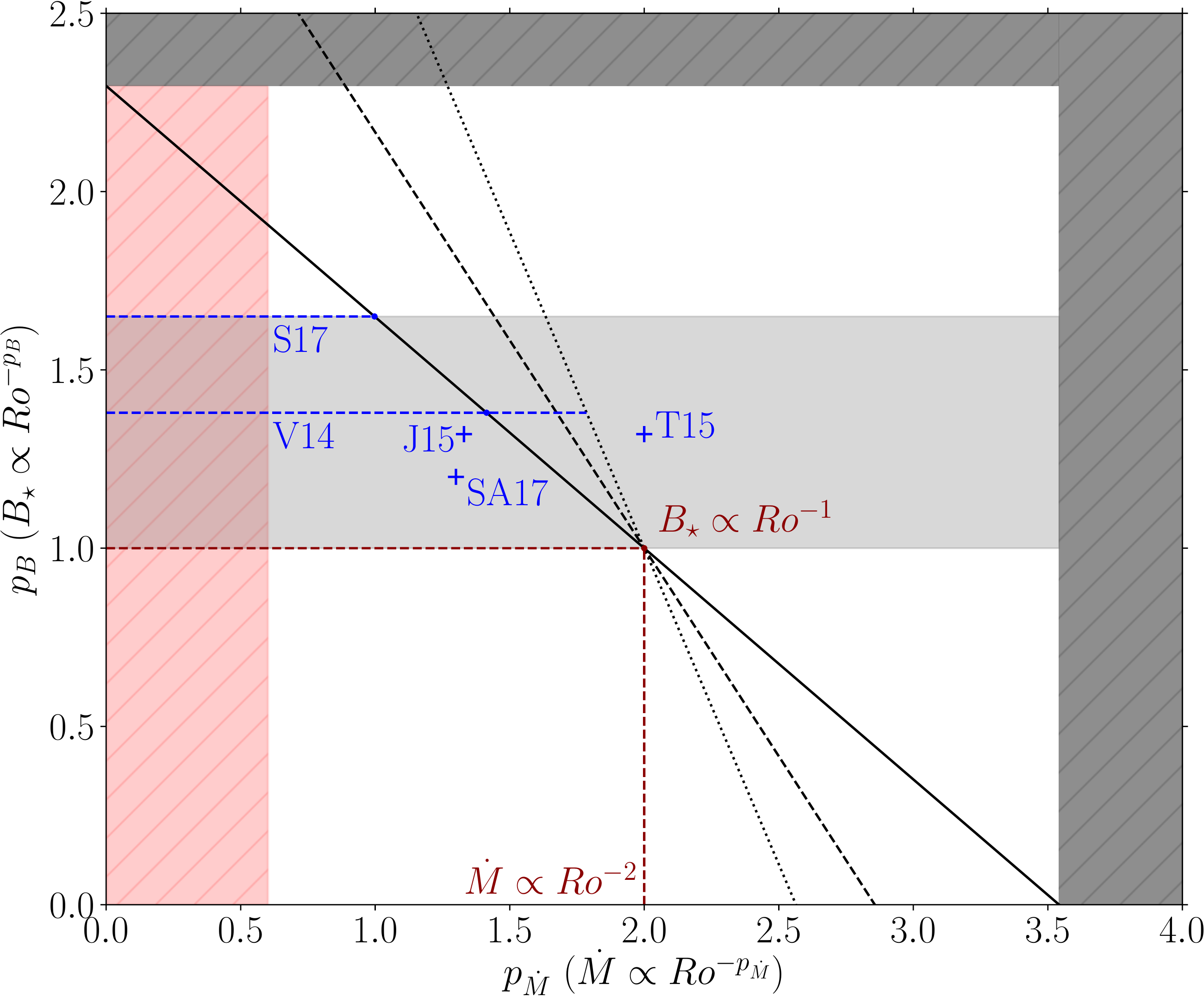}
\caption{Correlation \(p_B-p_{\dot M}\) for a dipolar (solid black line), a quadrupolar (dashed black line) and an octupolar (dotted black line) topology in the unsaturated regime. Dashed blue lines: magnetic field prescriptions from \citet{see17} [S17] and \citet{vidotto14} [V14]. Blue crosses: magnetic field and mass loss prescriptions from \citet{johnstone15b} [J15], \citet{tu15} [T15] and \citet{ardestani} [SA17]. In dark grey: excluded region in the dipolar case. Dashed dark red lines: prescriptions compatible with the Skumanich law for any magnetic topology. In light red: exponents corresponding to mass losses outside the envelope of \citet{wood05} data. The light grey band corresponds to the upper and lower bounds from ZDI statistical studies (see \S 3.4 for more details).}
\label{pBpM}
\end{figure}

In Figure \ref{pBpM} we illustrate the interdependencies in this first set of exponents. More precisely, the \(p_B\) exponent is expressed as a function of \(p_{\dot M}\) from equation \eqref{eqn:cons_rot1} for a dipolar field (solid black line), a quadrupolar field (dashed black line) and an octupolar field (dotted black line) in the case of an unsaturated rotation regime. The dashed blue lines correspond to magnetic field prescriptions from \citet{see17} (S17) and \citet{vidotto14} (V14). The blue crosses represent the magnetic field and mass loss prescriptions from \citet{johnstone15b,tu15,ardestani} (\textit{resp.} J15, T15, SA17). Those scaling laws have been derived to reproduce rotation rates of open clusters and are quite in agreement with a square root spin-down law. As already pointed out by \citet{skumanich19}, a wide range of exponents is admissible from the different prescriptions considered.

The magnetic topology has also a significant influence on the \(\dot M-B_\star\) prescriptions. Indeed, the value of the \(m\) exponent decreases with an increasing complexity of the topology, corresponding to higher-order multipoles \citep{reville15a}. We will take here \(m=0.15\) for a quadrupolar field (see the dashed black line in Figure \ref{pBpM}) and \(m=0.11\) for an octupolar field (see the dotted black line in Figure \ref{pBpM}). More complex magnetic fields lead to steeper slopes in Figure \ref{pBpM} and therefore to a less constrained magnetic field. The mass loss, for its part, tends to be proportional to \(Ro^{-2}\). Only one prescription of \(B_\star\) and \(\dot M\) is compatible with the Skumanich law for any magnetic topology (see the dark red dashed lines in Figure \ref{pBpM}) and corresponds to the one highlighted by \citet{skumanich19}. This configuration leads to a linear magnetic field-rotation relation and a quadratic mass loss-rotation relation. This way, such a prescription may be used as a first estimate of \(B_\star\) and \(\dot M\) based on spin-down considerations without having to assume a particular magnetic topology. However, in the following sections, we will keep unspecified prescriptions for the magnetic field and the mass-loss rate in order to study the influence of additional observational constraints. It is important to note that complex fields can significantly modify the wind braking torque itself \citep{reville15a,garraffo16}. Furthermore, from those considerations, \citet{garraffo18} were able to reproduce the bimodal distribution of slow and fast rotators.

Since the dipole component tends to dominate the wind braking torque for mixed geometries \citep{finleymatt18}, such a topology is assumed by default in the following sections.

\subsection{Constraints on the mass-loss rate}

Given the one-to-one correspondence between the \(B_\star\) and \(\dot M\) exponents, constraints on the mass loss will affect the magnetic field and reciprocally. First, the mass loss is enhanced by stellar rotation \citep{wood05,suzuki13,holzwarth}, leading to \(p_{\dot M} \geq 0\). Therefore, we obtain
\begin{equation}
p_B\leq \frac{1}{2m}\approx 2.3\qquad\text{(unsaturated)}
\end{equation}
\begin{equation}\label{eqn:pBsat}
p_B\leq 0\qquad\text{(saturated)}.
\end{equation}

Furthermore, \citet{wood02,wood05} showed a correlation between the mass loss and the X-ray stellar flux \(F_X\) for unsaturated main-sequence stars, expressed as
\begin{equation}\label{eqn:wood}
\dot M \propto R_\star^2 F_X^w,
\end{equation}
where \(w\) is a constant between \(w_\text{min} \approx 0.3\) and \(w_\text{max} \approx 1.9\), to be consistent with \citet{wood05} observations. To convert this correlation into a \(\dot M(Ro,R_\star,M_\star)\) prescription, we consider a relationship between the coronal activity and the stellar rotation, for unsaturated main-sequence stars:\\
\begin{equation}\label{eqn:RoLX}
\frac{L_X}{L_\star} \propto Ro^{-p_L},
\end{equation}
with \(L_X\) the X-ray stellar luminosity, \(L_\star\) the luminosity of the star and \(p_L\) an exponent between \(p_{L,\text{min}} = 2\) and \(p_{L,\text{max}} = 3\) \citep{pizzolato,wright,reiners}. If we take into account the scaling laws from Section 3.1, it is possible to estimate the mass loss as a function of the Rossby number and the stellar mass from stellar activity considerations. Indeed, the X-ray flux becomes
\begin{equation}\label{eqn:Fx}
F_X \propto Ro^{-p_L}M_\star^{\eta-2\xi},
\end{equation}
which leads to the mass loss
\begin{equation}
\dot M \propto Ro^{-p_L w} M_\star^{(\eta-2\xi)w+2\xi}.
\end{equation}
By identification, the exponents of \(\dot M\) can be inferred from the set \(\{\eta,\xi,p_L,w\}\):
\begin{equation}\label{eqn:conswood1}
p_{\dot M} = p_L w
\end{equation}
\begin{equation}\label{eqn:conswood2}
\xi r_{\dot M}+m_{\dot M}=(\eta-2\xi)w+2\xi.
\end{equation}
To be consistent with \citet{wood05} data, the most flexible constraint on the \(p_{\dot M}\) exponent can be obtained from equation \eqref{eqn:conswood1} as
\begin{equation}\label{eqn:wooddata}
0.6 \leq p_{\dot M} \leq 5.7.
\end{equation}
According to equation \eqref{eqn:wooddata}, the values of \(p_{\dot M}\) leading to \((F_X,\dot M)\) outside the envelope of \citet{wood05} data define the exclusion red regions in Figure \ref{pBpM} and 3. Equations \eqref{eqn:conswood1} and \eqref{eqn:conswood2} then give an additional constraint on the different exponents of the mass loss prescription
\begin{equation}
\xi r_{\dot M}+m_{\dot M}-2\xi = \frac{\eta-2\xi}{p_L}p_{\dot M}.
\end{equation}
This will dictate itself the mass and radius dependency of the stellar magnetic field, easier to observe, through equations \eqref{eqn:cons_rot1} to \eqref{eqn:cons_rot4}, such that
\begin{equation}\label{eqn:prmB}
4m(\xi r_B+m_B)=\xi(a-m-4)+b+m- \frac{\eta-2\xi}{p_L}\mathopen(2-4mp_B\mathclose).
\end{equation}

\begin{figure}[!h]
\centering
\includegraphics[scale=0.32]{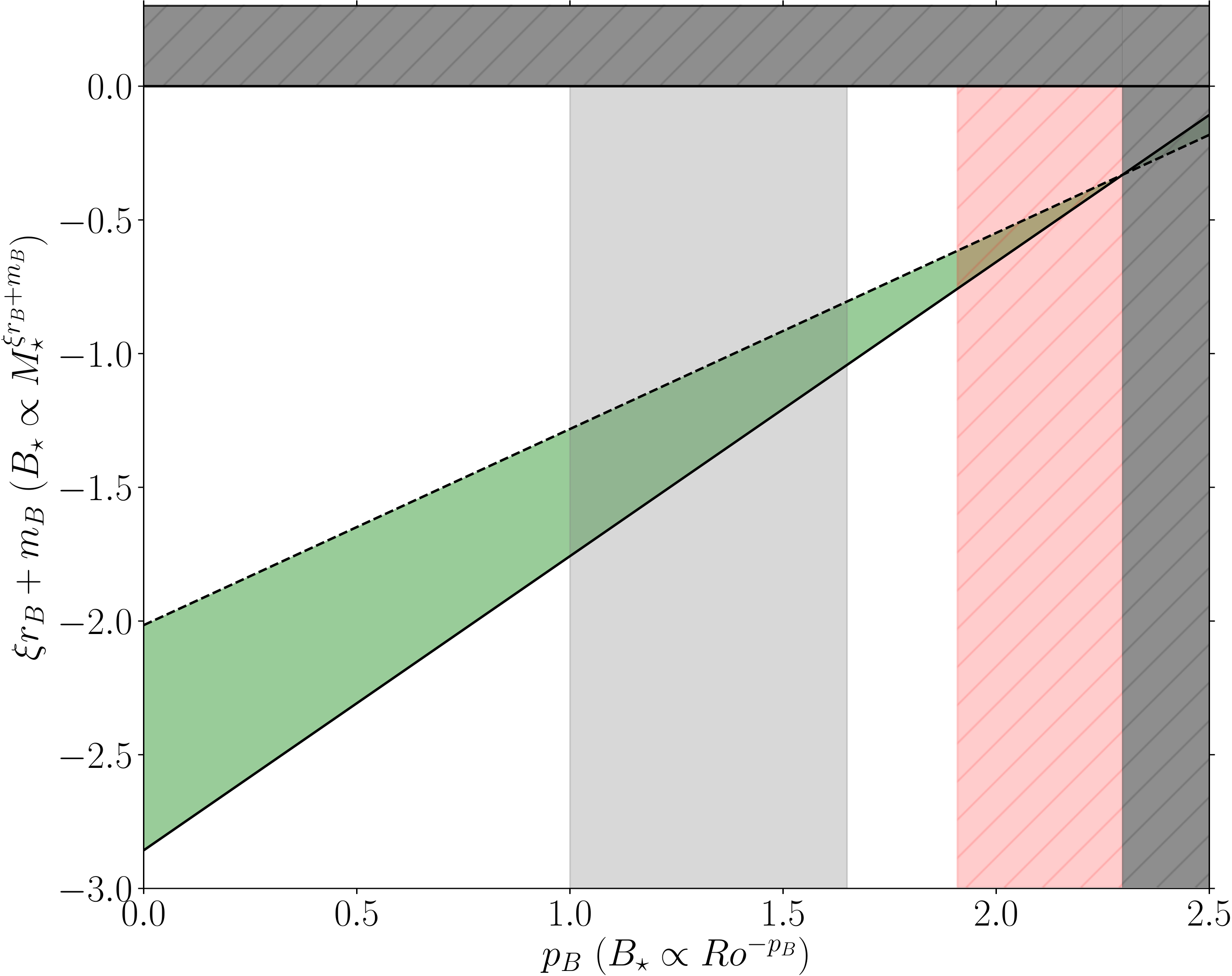}
\caption{Correlation between \(p_B\) and \(\xi r_B+m_B\) in the unsaturated regime for \(p_L = 2\) (black line) and \(p_L = 3\) (dotted). Here, \(\eta = 4\) and \(\xi = 0.9\). In dark grey: excluded region. In red: exponents corresponding to mass losses outside the envelope of \citet{wood05} data. In green: exponents consistent with the mass-loss constraints. The light grey band corresponds to the upper and lower bounds from ZDI statistical studies (see \S 3.4 for more details).}
\label{pBrmB}
\end{figure}
Equation \eqref{eqn:prmB} shows that the mass-radius dependency of the magnetic field can be inferred from its Rossby number dependency, as shown in Figure \ref{pBrmB}, where the green area represents the exponents of the magnetic field prescription compatible with the different constraints we considered. The negative values of \(\xi r_B+m_B\) show that the magnetic field should decrease with the stellar mass, which seems to be in agreement with dynamo models and observations \citep{johnskrull,brun15}. 

One can also notice in Figure \ref{pBrmB} that a stronger Rossby number dependency, for high values of \(p_B\), will lead to a weaker but non negligible dependency on stellar mass through the \(\dot M-F_X\) correlation. Futhermore, by considering the different exclusion regions (see the hatched areas in Figure \ref{pBrmB}), we can infer that a magnetic field which is solely a function of the Rossby number (\textit{i.e.} \(r_B = m_B = 0\)) cannot be consistent with the observed rotational dependency of \(L_X\) and \(\dot M\), as well as with the \(\dot M-F_X\) correlation. To fulfill this set of conditions, we need to add an explicit mass dependency to the stellar magnetic field.

Only a linear combination of \(r_B\ \&\ m_B\) is here constrained, because of the mass-radius relationship. However, it is possible to discriminate the mass and radius dependencies by considering the magnetic topology. 

\subsection{Constraints on the magnetic field}

The stellar magnetic field can be constrained in a variety of ways. Spectropolarimetric studies have exhibited correlations between the large scale magnetic field and other stellar parameters. For example, in \citet{vidotto14}, \(B_\star\) is shown to scale as \(Ro^{-1.38\pm 0.14}\), which leads to a mass loss \(p_{\dot M} = -1.41\pm 0.22\). The X-ray luminosity of the star can also be considered to infer the magnetic field. We assume here a correlation \(L_X-B_\star\) such that \citep{petsov,vidotto14}
\begin{equation}
\frac{L_X}{L_\star}\propto B_\star^v,
\end{equation}
which leads to \(p_B = p_L/v \). Assuming for instance \(p_B = -1.38 \pm 0.14\) and \(p_L = 2.5 \pm 0.5\) leads to \(v = 1.81 \pm 0.6\), to be compared with the value \(v = 1.61\pm 0.15\) obtained in \citet{vidotto14} through a power-law fit.

Since the magnetic field is enhanced by the stellar rotation \citep{noyes,brandenburg,petit08}, the exponent \(p_B\) has to be nonnegative, leading to the following constraint on the mass-loss rate in the unsaturated regime
\begin{equation}
p_{\dot M}\leq \frac{2}{1-2m}\approx 3.5.\qquad\text{(unsaturated)}
\end{equation}
At saturation, with equation \eqref{eqn:pBsat}, such a condition gives \(p_B = 0\ \&\ p_{\dot M} = 0\), which means a mass loss and a magnetic field independent of the stellar rotation rate.

We will consider in the following sections lower and upper bounds for the values of \(p_B\) and \(p_{\dot M}\), in order to be consistent with observational trends. Measured stellar magnetic field, from Zeeman broadening and ZDI studies \citep{montesinos,vidotto14,see17}, only exhibited linear or super-linear dependencies between the large-scale magnetic field and the Rossby number. We will then take \(p_B \geq 1\) to take this fact into account. The \citet{see17} prescription, \textit{i.e.} \(p_B = 1.65\), will be used as an upper bound. The values of all the exponents associated to these two scenarii are given in Table \ref{tab:typical}. It is a common knowledge that ZDI maps, representing the large-scale unsigned 
magnetic flux, do not give any information about the small-scale magnetic field, that might be dominant for young fast rotating stars. However, a correlation between measurements from Zeeman-Doppler Imaging and Zeeman Broadening \citep{see19} gives some confidence in the general trends found in the literature.

The bounds we assumed above are greater than the exponents predicted by some scaling laws for stellar dynamos  \citep[\(p_B = 0\) for the equipartition, \(p_B = 1/4\) for the buoyancy regime and \(p_B = 1/2\) for the magnetostrophy regime ; for a review, see][]{augustson17a}. This discrepancy likely comes from the fact that dynamo scaling laws and ZDI observations do not relate to the same magnetic field. The measured magnetic field from ZDI studies corresponds to the average unsigned photospheric flux \(\left<B_V\right>\), which is an estimate of the large scale magnetic field at the stellar surface. Dynamo-based scaling laws aim to estimate the stellar magnetic field over a wide spectral range. The latter can be linked to the average unsigned surface field strength, \(\left<B_I\right>\), obtained from Zeeman broadening, by means of a filling factor \(f\) representing the fraction of the stellar surface which is magnetized (see \citet{reiners12} for more details). \citet{see19}, by exhibiting a correlation between \(\left<B_V\right>\) and \(\left<B_I\right>\), estimated a filling factor from a dynamo-produced magnetic field in the equipartition regime and showed a strong Rossby number dependency of their \(f\) estimate, which could be an avenue towards the explanation of this difference.

In a nutshell, it is possible by relying on a wind torque parametrization to provide \(B_\star\) and \(\dot M\) estimates consistent with spin-down and X-ray emission constraints (cf. equations \eqref{eqn:cons_rot1} to \eqref{eqn:cons_rot4} and equation \eqref{eqn:prmB}), given for example the \(B_\star-Ro\) relation (\textit{i.e.} the \(p_B\) exponent). In a quest to better understand the link between stellar and wind properties, we will expand our set of prescriptions to probe the coronal properties of the star by relying upon a wind model.

\subsection{Constraints on the coronal properties}
\subsubsection{Probing the coronal properties: the role of magnetic topology}
As seen in the previous section, the mass loss behavior can be constrained by the stellar magnetic field and the wind braking torque. This way, the knowledge of the \(\dot M\) expression will allow us to go further by inferring prescriptions for the coronal temperature and density, by means of the X-ray stellar emission. Indeed, their rotational dependency have often been constrained in the literature through the high-energy activity of the star \citep{mestel87,ivanova,holzwarth}.

\begin{figure}[!h]
\centering
\includegraphics[scale=0.29]{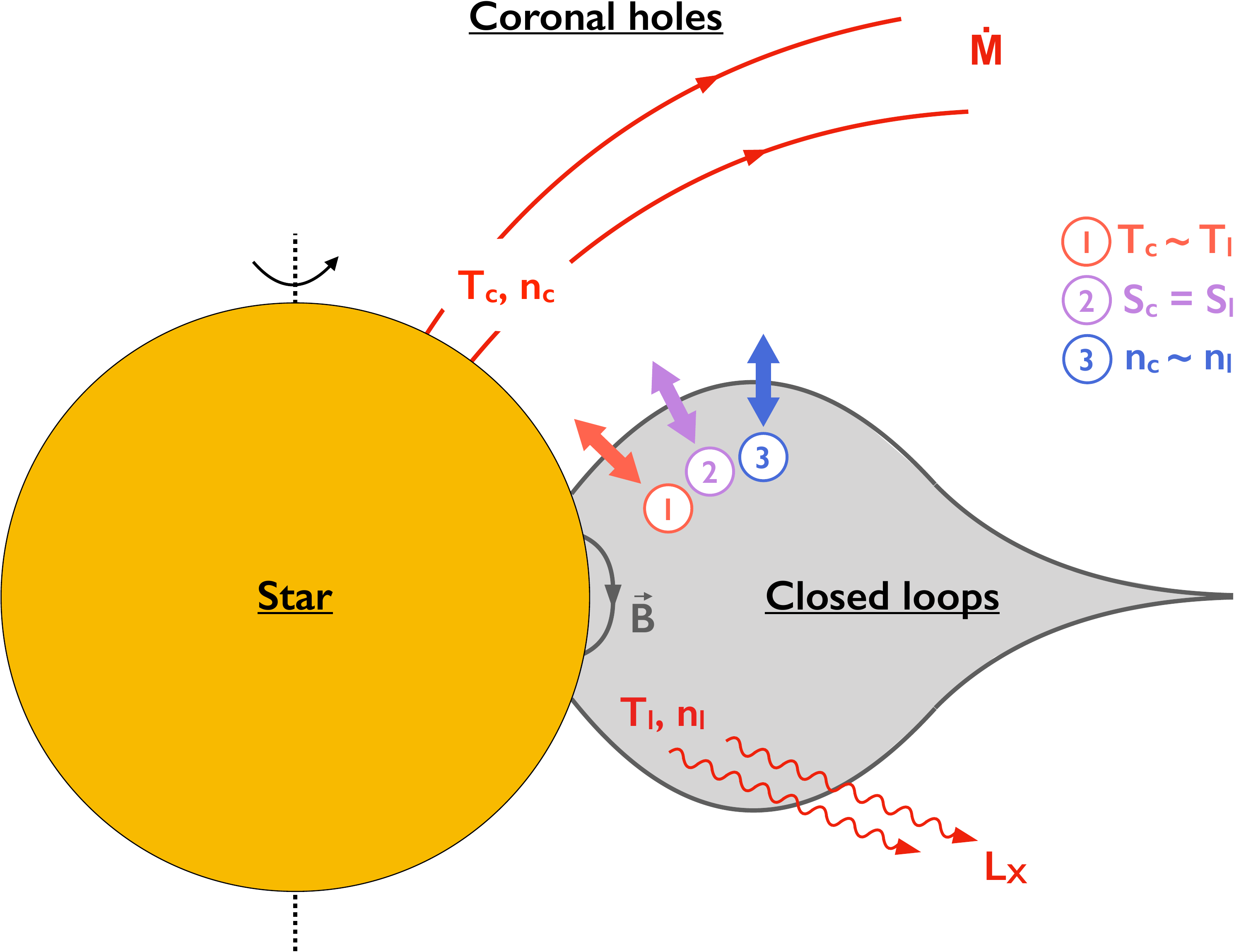}
\caption{Sketch of the coupling between the coronal temperature and density in the open-field and the closed-field regions, the X-ray luminosity and the mass-loss rate. Several scenarii are considered to connect the closed loops to the coronal holes. Scenario 1, in red: single temperature scaling. Scenario 2, in purple: entropy equilibrium. Scenario 3, in blue: single density scaling.}
\label{scenarii}
\end{figure}

However, X-ray emission and wind acceleration (at least for the fast wind) are believed to arise from different regions in the corona (from dead zones and coronal holes respectively). Therefore, we will here assume that the coronal temperature \(T_c\) and the base density \(n_c\) in the open-field regions are the quantities ruling the mass loss for a given wind model (cf. equation \eqref{eqn:mdotwindfin} for example), whereas the X-ray luminosity can be inferred from radiative losses by knowing the temperature \(T_l\) and the density \(n_l\) in closed-field regions (cf. Appendix B).
In other words, the \(\dot M\) prescription from the previous sections allow us to link \(T_c\) and \(n_c\), while the Rossby dependency of \(L_X\), which has to be consistent with equation \eqref{eqn:RoLX}, correlates \(T_l\) and \(n_l\). All those connexions are summarized in Figure \ref{scenarii}. Since \(T_l\) can be obtained from observations \citep{preibisch97,johnstonegudel,wood18}, we need an additional constraint to relate (\(T_c,\ n_c\)) to (\(T_l,\ n_l\)).\\

To this end, we will consider three scenarii to couple the open-field and the closed-field regions:

\begin{itemize}
    \item \textit{Scenario 1} (in red in Figure \ref{scenarii}): we consider a common scaling law for \(T_c\) and \(T_l\) \citep{johnstone15a,ofionnagain} which leads to different scalings for the densities.\\
    
    \item \textit{Scenario 2} (in purple in Figure \ref{scenarii}): we assume an entropy equilibrium between the closed loops and the coronal holes.\\
        
    \item \textit{Scenario 3} (in blue in Figure \ref{scenarii}): we assume a common scaling for \(n_c\) and \(n_l\) \citep{ivanova,holzwarth,see14} which leads to different scalings for the temperatures.
\end{itemize}

It is worth noticing that scenarii 1 and 3 lead to drastically different trends for the temperature \(T_c\). Therefore, a scenario where both the density and the temperature possess the same scaling in the open and closed field regions is unrealistic.

Such an approach allows us to deal with the local magnetic field distribution in an admittedly simplified way, by distinguishing quiet open-field regions from closed loops associated to active regions. This distinction translates into different coronal temperature and density prescriptions, which are always chosen to be compatible with our derived wind mass-loss rate in the polytropic formalism. Of course at the surface of the star the plasma dynamics and heating mechanisms are much more involved than in our simplified approach \citep{wedemeyer}.  Still this is a first step to characterize the general properties of the coronae of cool stars, and we intend in the future to consider more realistic modeling for the detailed coronal heating mechanism.

In what follows, we will see how to constrain the coronal temperatures and densities from \(L_X\) and \(\dot M\) (Sections 3.5.2 and 3.5.3 respectively). Then, we will inventory the different scenarii in Sections 3.5.4, 3.5.5 and 3.5.6.

\subsubsection{Coronal temperature and density: X-ray luminosity consistency}
The temperature and the density in closed loops are connected to the X-ray luminosity, which has to be consistent with the prescriptions we already adopted (cf. equation \eqref{eqn:RoLX}). From radiative losses considerations, the X-ray luminosity can be expressed as (cf. Appendix B)
\begin{equation}\label{eqn:Lxrss}
L_X\propto \left(\frac{R_\star}{R_\odot}\right)^3\left(\frac{B_\star}{B_\odot}\right)\left(\frac{n_l}{n_{l,\odot}}\right)^{\frac{3}{2}}\left(\frac{T_l}{T_{l,\odot}}\right)^{-\frac{7}{6}}.
\end{equation}

In order to standardize the different prescriptions, we will consider a power-law expression for the temperature and the density in the dead zones, \textit{i.e.}
\begin{equation}\label{eqn:Tlpres}
T_l \propto \left(\frac{Ro}{Ro_\odot}\right)^{-p_{T,l}}\left(\frac{R_\star}{R_\odot}\right)^{r_{T,l}} \left(\frac{M_\star}{M_\odot}\right)^{m_{T,l}},
\end{equation}
\begin{equation}\label{eqn:nlpres}
n_l \propto \left(\frac{Ro}{Ro_\odot}\right)^{-p_{n,l}}\left(\frac{R_\star}{R_\odot}\right)^{r_{n,l}} \left(\frac{M_\star}{M_\odot}\right)^{m_{n,l}}.  
\end{equation}
Furthermore, for main-sequence stars in the unsaturated regime, we will assume a correlation between the X-ray flux and the coronal temperature in the closed loops such that \citep{preibisch97,johnstonegudel, wood18}
\begin{equation}\label{eqn:TlFx}
T_l \propto F_X^{j_l},
\end{equation}
with an exponent \(j_l=0.26\) \citep{johnstonegudel}. Equation \eqref{eqn:Fx}, with relationships from Section 3.1, then result in a \(T_l\ (Ro, M_\star)\) formulation
\begin{equation}\label{eqn:TlRoMR}
T_l \propto Ro^{-p_L j_l}M_\star^{(\eta-2\xi)j_l}. 
\end{equation}
This leads by identification to
\begin{equation}\label{eqn:pTL}
p_{T,l} = p_L j_l.
\end{equation}
The rotational dependency of the density can therefore be deduced from the magnetic field prescription for a given set \(\{p_L,j_l\}\) by ensuring the consistency with equation \eqref{eqn:RoLX}:
\begin{equation}\label{eqn:pn}
p_{n,l} = \frac{2}{3}(p_L - p_B) +\frac{7}{9}p_L j_l.
\end{equation}
The X-ray luminosity therefore gives us the possibility to constrain the Rossby dependency of the temperature and the density in closed loops.

We now need to study the connection between the X-ray luminosity and the temperature in coronal holes, which will allow us to infer the mass-radius dependency of \(T_c\) from its rotational dependency, as we did for the mass-loss rate in Section 3.3. X-ray emission and wind acceleration have similar sources closely linked to the heating of the corona, which is probably due to the transport of energy from the photosphere through weakly dissipative Alfvén waves. These phenomena involve steep density gradients \citep{heyvaerts} and nonlinear interactions between inward and outward perturbations \citep{velli89}, among others \citep[for a review, see][]{mathioudakis,cranmer17}. While this process is efficient in closed loops thanks to the magnetic topology, it requires wave reflections \textit{e.g.} through the parametric instabilities in coronal holes \citep{reville18}, reducing the heating efficiency in those regions. To take this behavior into account, we will assume a correlation between the X-ray flux and the coronal temperature \(T_c\) like in the closed loops case, with an unspecified exponent, different from \(j_l\) because of the possible difference of heating efficiency between open-field and closed-field regions. 

\subsubsection{Coronal temperature and density: polytropic wind model considerations}
The temperature and the density in coronal holes need to be consistent with our mass loss prescription, constrained in Sections 3.1 to 3.4. We will consider the following power-law expression for the coronal temperature:
\begin{equation}\label{eqn:Tcpres}
T_c \propto \left(\frac{Ro}{Ro_\odot}\right)^{-p_T}\left(\frac{R_\star}{R_\odot}\right)^{r_T} \left(\frac{M_\star}{M_\odot}\right)^{m_T}.
\end{equation}
Such a prescription has to fulfill the condition \(\dot M >0\), which defines a maximal value for the \(p_T\) exponent (see Appendix C for more details).

Knowing the value of \(T_c\) for a given wind model, it is possible to infer the coronal density to obtain a consistent mass-loss rate. As we saw in Section 2.4 with the equation \eqref{eqn:mdotwindfin}, the mass loss can be expressed as a function of stellar parameters and coronal properties. To be consistent with the power-law prescription from equation \eqref{eqn:mdot}, the coronal density \(n_c\) has to be expressed as
\begin{equation} \label{eqn:ncwind}
n_c \propto Ro^{-p_{\dot M}}R_\star^{r_{\dot M}}M_\star^{m_{\dot M}-2} T_c^{\frac{3}{2}}\left[1-\frac{T_{\text{min},\odot}}{T_\odot}\frac{M_\star}{M_\odot}\frac{R_\odot}{R_\star}\frac{T_\odot}{T_c}\right]^{\frac{3\gamma-5}{2(\gamma-1)}} .
\end{equation}
It is important to keep in mind that this equation is valid if \(c_s/v_\text{esc}\ll 1\), where \(c_s\) is the speed of sound in the stellar corona, and \(v_\text{esc}\) the escape velocity at the stellar surface (cf. Appendix A). In the light of this condition, high coronal temperatures could invalidate this analytical expression of \(n_c\).

If we assume that the coronal temperature varies slightly for the stellar parameters we consider, we can approximate this expression to a power law on the Main Sequence such that (see details in Appendix D)
\begin{equation}\label{eqn:ncpowerlaw}
n_c \propto \left(\frac{Ro}{Ro_\odot}\right)^{-p_n}\left(\frac{R_\star}{R_\odot}\right)^{r_n}\left(\frac{M_\star}{M_\odot}\right)^{m_n},
\end{equation}
with:
\begin{equation}\label{eqn:pnpT}
p_n = p_{\dot M}+\left(\frac{3}{2}-F(\gamma)\right)p_T,
\end{equation}
\begin{equation}\label{eqn:rnrT}
r_n = r_{\dot M}+\frac{3}{2} r_T-F(\gamma)(1+r_T),
\end{equation}
\begin{equation}\label{eqn:mnmT}
m_n = m_{\dot M}-2+\frac{3}{2} m_T-F(\gamma)(m_T-1)
\end{equation}
\begin{equation}
\text{and } F(\gamma)= \frac{5-3\gamma}{2(\gamma-1)}\frac{T_{\text{min},\odot}/T_\odot}{1-T_{\text{min},\odot}/T_\odot}.
\end{equation}

Therefore, for a given mass-loss rate prescription, a one-to-one correspondence between \(\{p_T,r_T,m_T\}\) and \(\{p_n,r_n,m_n\}\) occurs by considering a pressure-driven polytropic wind. Furthermore, we are able to infer \(\xi r_T + m_T\) from the \(p_T\) exponent thanks to a \(T_c-F_X\) correlation (cf. Appendix C). Given that the Rossby dependency of \(T_l\) and \(n_l\) is already known through equations \eqref{eqn:pTL} and \eqref{eqn:pn} respectively, we only need one additional constrain to fully determine the expression of \(T_c\) and \(n_c\). We choose here to connect the rotational dependency of the temperature and the density in open and closed regions by means of the three scenarii we presented in Section 3.5.1. We detail now the implications of these scenarii in sections 3.5.4, 3.5.5 and 3.5.6. 

\subsubsection{Scenario 1: single temperature scaling}
To connect closed loops to coronal holes, we can adopt a single scaling law for the temperature (\textit{i.e.} \(T_l \propto T_c\)), leading to a same X-ray flux-temperature correlation in both regions. This way, the \citet{johnstonegudel} prescription, coupled with the \(L_X-Ro\) relation and the results from Section 3.1, provides a complete expression for \(T_c\). One can not assume weak variations of the coronal temperature, which means that a power-law expression for \(n_c\) may be a loose approximation in this scenario (cf. Appendix D). However, the coronal density can be inferred directly from our wind model through equation \eqref{eqn:ncwind}, in order to keep a consistent mass loss. Those considerations then allow us to estimate \(T_c\) and \(n_c\) by means of the following prescriptions

\begin{equation}\label{eqn:pT_isoT}
p_T = p_L j_l,
\end{equation}
\begin{equation}\label{eqn:rmT_isoT}
\xi r_T + m_T = \frac{\eta - 2\xi}{p_L}p_T,
\end{equation}
\begin{equation}\label{eqn:nc_isoT}
n_c \propto Ro^{-p_{\dot M}}R_\star^{r_{\dot M}}M_\star^{m_{\dot M}-2} T_c^{\frac{3}{2}}\left[1-\frac{T_{\text{min},\odot}}{T_c}\frac{M_\star}{M_\odot}\frac{R_\odot}{R_\star}\right]^{\frac{3\gamma-5}{2(\gamma-1)}} .
\end{equation}
In a saturated rotation regime, it is impossible from equation \eqref{eqn:nc_isoT} to keep a constant value for both \(T_c\) and \(n_c\). We will assume by simplicity a single temperature at saturation, the corresponding coronal density being inferred from the \(\dot M\) prescription.\\

The robustness of the single temperature scaling hypothesis can nevertheless be questioned. In closed loops, available estimates of \(T_l\) \citep{preibisch97,johnstonegudel,wood18} fall back on an emission measure weighted average coronal temperature, based on heavy ions emission \citep{gudel07}. From the wind model point of view, based on the modelling of a perfectly ionized hydrogen gas, the electron temperature (which is in this context similar to the proton temperature) is required to compute the mass-loss rate. Thus, a single scaling law for both temperatures may appear as a loose assumption. Indeed, if different radial profiles of temperature are observed in the solar wind for these two populations \citep{cranmer17}, with a higher temperature for heavy ions, their dependency on stellar parameters is still unknown.

Furthermore, if we assume \(T_c \propto T_l\), slower rotators may have a colder corona, leading to extremely strong densities to keep a consistent mass loss. As an example, for a solar twin with a rotation period of 56 days, the single temperature scaling scenario leads to \(n_c \approx 19.58\ n_\odot\), which has to be compared to \(n_l \approx 0.48\ n_\odot\), according to equation \eqref{eqn:pn}. Coronal holes of slow rotators would be abnormally dense, which may suggest that a single scaling law for the temperature might be inconsistent with the other hypothesis of our formalism, especially the choice of our wind model. Because of those points we will consider other scenarii in what follows.

\subsubsection{Scenario 2: entropy equilibrium}
A second possibility is to assume entropy equilibrium between closed and open field regions allowing both density and temperature to vary simultaneously. This allows us to derive the following relation:
\begin{equation}
\ln T_l - (\gamma_{ad} - 1) \ln n_l = \ln T_c - (\gamma_{ad} - 1) \ln n_c,
\end{equation}
where \(\gamma_{ad} = c_p/c_v = 5/3\) is the standard adiabatic exponent for an ideal gas. If we assume a power-law expression for \(n_c\) as in equation \eqref{eqn:ncpowerlaw}, this balance then gives for the Rossby dependency of the temperatures and densities:
\begin{equation}
p_{T,l}-(\gamma_\text{ad}-1)p_{n,l} = p_T-(\gamma_\text{ad}-1)p_n.
\end{equation}
Along with equation \eqref{eqn:pnpT} ruling the rotational dependency of the mass-loss rate in our wind model, one can express the \(p_T\) and \(p_n\) exponents as follows:
\begin{equation}\label{eqn:pT_isoS}
p_T =\frac{(\gamma_\text{ad}-1)(p_{n,l}-p_{\dot M})-p_L j_l}{(\gamma_\text{ad}-1)[3/2-F(\gamma)]-1},
\end{equation}
\begin{equation}\label{eqn:pn_isoS}
p_n = p_{\dot M}+\left(\frac{3}{2}-F(\gamma)\right)p_T.
\end{equation}

The \(T_c-F_X\) correlation (cf. Appendix C), along with the wind model through equations \eqref{eqn:rnrT} and \eqref{eqn:mnmT}, dictate the mass-radius dependency of \(T_c\) and \(n_c\) as
\begin{equation}\label{eqn:rmT_isoS}
\xi r_T + m_T = \frac{\eta - 2\xi}{p_L}p_T
\end{equation}
\begin{equation}\label{eqn:rmn_isoS}
\xi r_n + m_n = \xi r_{\dot M} + m_{\dot M} - 2 + (\xi r_T + m_T) \left[\frac{3}{2}- F(\gamma)\right]+(1-\xi)F(\gamma)
\end{equation}
Equations \eqref{eqn:pT_isoS} to \eqref{eqn:rmn_isoS} then define the power-law expressions of \(T_c\) and \(n_c\) for this scenario.\\

At high rotation rates, we have seen in Section 3.3 that the linear saturation of the wind braking torque leads to a magnetic field and a mass loss independent of the Rossby number, \textit{i.e.} \(p_B = p_{\dot M} = 0\). A similar behavior has been observed for the X-ray activity of stars \citep{pizzolato,wright}. We assume here for simplicity that \(p_L = 0\) in the saturated regime. We can show by relying on calculations similar to those presented in Sections 3.5.2 and 3.5.3 that \(p_n = p_T = 0\) is an acceptable choice if we assume a similar behavior in closed loops.

\subsubsection{Scenario 3: single density scaling}

Let us finally consider that the coronal density has a similar behavior in the closed-field and open-field regions. This way, two different scaling laws will arise for the temperatures in closed or open-field regions. 
In this case, the coronal density \(n_c\), launching the wind in the open-field regions, is proportional to the density in closed loops \(n_l\), and therefore can be expressed as a power law with a Rossby exponent given by equation \eqref{eqn:pn} to ensure a consistent X-ray luminosity. Considering a power-law expression for \(n_c\) is a reasonable assumption if the temperature in coronal holes presents only small variations with respect to stellar parameters (cf. Appendix D). Such a supposition seems to be consistent with \citet{suzuki13}, who performed simulations of flux tubes heated by Alfvén wave dissipation in coronal holes. Indeed, they predicted a weak dependency of the coronal temperature on the stellar magnetic field, which can be extrapolated to more fundamental stellar parameters through a dynamo relationship (see equation \eqref{eqn:magfield}). We can now infer the expression of the Rossby-dependency of the coronal temperature (\textit{i.e.} the \(p_T\) exponent) through equation \eqref{eqn:pnpT}. All those aspects allow us to determine the rotational dependency of \(T_c\) and \(n_c\) thanks to the following prescriptions
\begin{equation}\label{eqn:pn_ison}
p_n = \frac{2}{3}(p_L - p_B) +\frac{7}{9}p_L j_l,
\end{equation}
\begin{equation}\label{eqn:pT_ison}
p_T = \frac{p_n - p_{\dot M}}{\frac{3}{2}-F(\gamma)}.
\end{equation}

As for the entropy equilibrium scenario, the mass-radius dependency of \(T_c\) and \(n_c\) is determined through the wind model and the \(T_c-F_X\) correlation (cf. Appendix C):
\begin{equation}\label{eqn:rmT_ison}
\xi r_T + m_T = \frac{\eta - 2\xi}{p_L}p_T,
\end{equation}
\begin{equation}\label{eqn:rmn_ison}
\xi r_n + m_n = \xi r_{\dot M} + m_{\dot M} - 2 + (\xi r_T + m_T) \left[\frac{3}{2}- F(\gamma)\right]+(1-\xi)F(\gamma).
\end{equation}
Equations \eqref{eqn:pn_ison} to \eqref{eqn:rmn_ison} therefore give us the possibility to estimate the coronal temperature and density in open regions.\\
\begin{figure}[!h]
\centering
\includegraphics[scale=0.39]{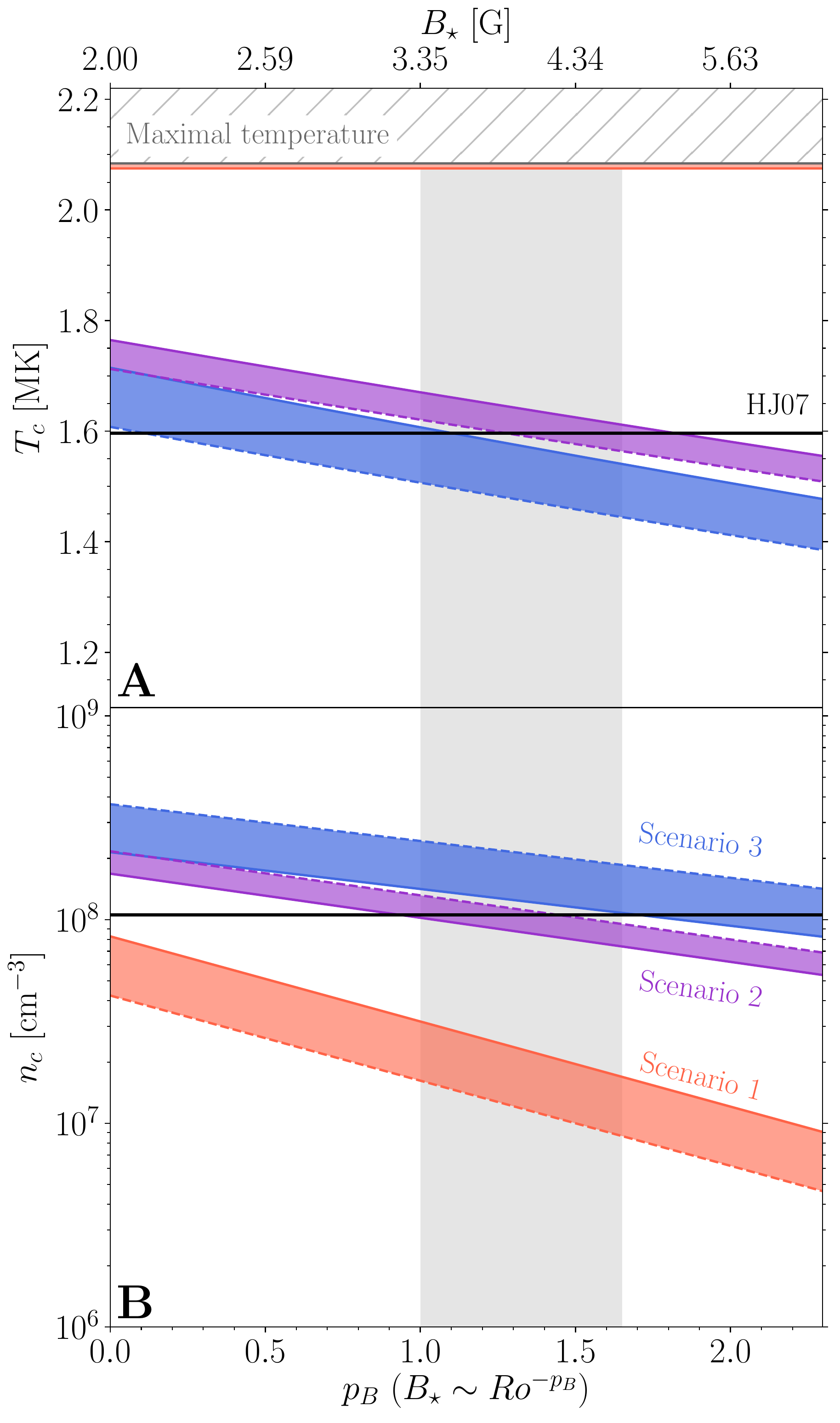}
\caption{Evolution of \(T_c\) (panel A) and \(n_c\) (panel B) as a function of \(p_B\), for a solar twin with a rotation period of 15 days. Solid line: \(p_L = 2\). Dashed lines: \(p_L = 3\). In black: \citet{holzwarth} prescription. In light red: single temperature scaling scenario. In purple: entropy equilibrium scenario. In blue: single density scaling scenario. In hatched grey: values of coronal temperatures leading to \(\dot M <0\). The light grey band corresponds to the upper and lower bounds of \(p_B\) we consider to be consistent with ZDI statistical studies.}
\label{solartwin}
\end{figure}

Like in Section 3.5.5, the assumption of a coronal temperature and density independent of the Rossby number in the saturated rotation regime is consistent with the different observational constraints we imposed.\\

Figure \ref{solartwin} shows the \(T_c\) and \(n_c\) estimates for a solar twin with a rotation period of 15 days (corresponding to \(Ro = 0.6\) in the \citet{ardestani} prescription) in the three physical scenarii we analyzed. We set \(\eta = 4,\ \xi = 0.9\) \citep{kippenhahn}, \(\gamma =1.05\ \text{and } T_\odot = 1.5\) MK \citep{reville16}. One can notice that a stronger dependency of the magnetic field on the Rossby number, \textit{i.e.} higher values of \(p_B\), leads to weaker values of \(T_c\  \text{and}\ n_c\) (apart from the coronal temperature in Scenario 1 which remains constant; see the light red band in panel A of Figure \ref{solartwin}). Furthermore, coronal temperatures obtained in scenarii 2 and 3 (see the blue and purple bands in panel A of Figure \ref{solartwin}) only vary a little around the solar value, which is consistent with the assumptions we made to express \(n_c\) as a power law. Since the star we consider here is a solar twin rotating more rapidly (\textit{i.e.} with \(Ro/Ro_\odot <1\)), the positiveness criterion of the mass-loss rate (cf. Appendix C) sets a maximal coronal temperature (see the dark grey line in panel A of Figure \ref{solartwin}). From this condition, one can affirm that scenarii 2 and 3 are compatible with a transsonic wind. However, in scenario 1, the condition \(\dot M > 0\) is only ensured for \(p_L \approx 2\), the coronal temperature being too high otherwise given a reasonable choice of (\(\gamma,T_\odot\)).  

Considering our lower and upper bounds for \(p_B\) as described in Section 3.4, single density scaling and entropy equilibrium lead to coronal temperatures close to the \citet{holzwarth} prescription (see the black horizontal line in Figure \ref{solartwin}). Scenario 3 leads to temperatures weaker than the predicted value of \citet{holzwarth}, resulting in slightly higher densities for a given mass-loss rate (see the blue band in panel B of Figure \ref{solartwin}). On the contrary, marginally higher temperatures and lower densities can be observed in scenario 2 (see the purple band in panel B of Figure \ref{solartwin}). Scenario 3, with a \(T_c\) prescription independent of the magnetic field and the mass-loss scaling laws, leads up to the highest values of the coronal temperature and the lowest values of the coronal density shown in Figure \ref{solartwin} (in light red in Figure \ref{solartwin}).

In the light of those different aspects, scenarii 2 and 3 seem more likely to account for a consistent stellar spin-down with a polytropic pressure-driven wind.

\section{Scaling laws and observations of individual systems}

\subsection{Using our scaling laws: a practical guide}

\subsubsection{Rossby number and rotation regime}

To deal with individual systems, we need to know their Rossby number to rely on our prescriptions. We will here use for the sake of simplicity the stellar Rossby number (cf. equation \eqref{eqn:def_rossby}) and more precisely the \citet{ardestani} prescription for the convective turnover time. Following their approach, this characteristic time is defined as the ratio between the pressure scale height and the convective velocity estimated with the mixing length theory. They computed the relevant quantities at half a pressure scale height over the base of the convective zone and followed their evolution by means of the CESAM stellar evolution code \citep{morel}. The formulation they obtained has the advantage of being valid during the Pre-Main Sequence and the Main Sequence for metallicities ranging from [Fe/H] = -0.5 to 0.5, by falling back on the stellar convective mass as the control parameter. Given the age of the system we will consider, such an expression can be simplified for main-sequence stars to depend only on the stellar mass and the stellar radius, which leads to the following formulation for the convective turnover time:
\begin{equation}
\tau_c \propto M_\star^{-1}R_\star^{-1.2}. 
\end{equation}
This prescription leads to a solar value \(Ro_\odot = 1.113\) and a saturation value \(Ro_\text{sat} = 0.09\).

\subsubsection{Scaling laws: numerical values of the exponents}

In the previous sections, we have been able to constrain the magnetic field, the mass loss, the coronal temperature and the coronal density from wind braking considerations, by assuming the following prescriptions (at least in the entropy equilibrium and the single density scaling scenarii):
\begin{equation}
\Gamma_\text{wind} \propto \dot{M}^{1-2m}B_\star^{4m} R_\star^{2+5m}M_\star^{-m}\Omega_\star\left[1+\frac{f^2}{K^2}\right]^{-m}
\end{equation}
\begin{equation}
B_\star\propto Ro^{-p_B}\left(\frac{R_\star}{R_\odot}\right)^{r_B}\left(\frac{M_\star}{M_\odot}\right)^{m_B}
\end{equation}
\begin{equation}
\dot M\propto Ro^{-p_{\dot M}}\left(\frac{R_\star}{R_\odot}\right)^{r_{\dot M}}\left(\frac{M_\star}{M_\odot}\right)^{m_{\dot M}} 
\end{equation}
\begin{equation}
T_c \propto Ro^{-p_T}\left(\frac{R_\star}{R_\odot}\right)^{r_T} \left(\frac{M_\star}{M_\odot}\right)^{m_T} 
\end{equation}
\begin{equation}
n_c \propto Ro^{-p_n}\left(\frac{R_\star}{R_\odot}\right)^{r_n} \left(\frac{M_\star}{M_\odot}\right)^{m_n} 
\end{equation}

\begin{table}[!h]
      \centering
      \caption{\label{tab:typical} Parameters defining the \(\Gamma_\text{wind},\ B_\star,\ \dot M,\ T_c\ \&\ n_c\) prescriptions for \(p_B =1\) and \(p_B = 1.65\) in the single density scaling scenario.}
         \begin{tabular}{llc}
            \hline
            \noalign{\smallskip}
            \text{Lower bound} & \text{Upper bound} & \text{Equation}\\
            \noalign{\smallskip}
            \hline
            \noalign{\smallskip}
            \textit{Free parameters.}\\
            \noalign{\smallskip}
            $\gamma$ = 1.05\tablefootmark{a}& - & \eqref{eqn:mdotwindfin}\\
            $m = 0.2177\tablefootmark{b}$ & - & \eqref{eqn:torque}\\
            \noalign{\smallskip}
            \hline
            \noalign{\smallskip}
            \textit{Constrained parameters.}\\
            \noalign{\smallskip}
            $T_\odot = 1.5$ MK \tablefootmark{a} & - & \eqref{eqn:mdotwindfin}\\
            $\eta = 4\tablefootmark{c},\ \xi=0.9\tablefootmark{c}$ & - & \eqref{eqn:LM}, \eqref{eqn:RM} \\
            $p_L = 2\tablefootmark{d}$ & - &\eqref{eqn:RoLX}\\
            \\
            $a=3.1\tablefootmark{e},\ b=0.5\tablefootmark{e}$ & - &\eqref{eqn:matt15unsat}\\
            \\
            $p_B = 1$ & $p_B = 1.65$\tablefootmark{f}& \eqref{eqn:magfield}\\
            $\xi r_B+m_B = -1.76$ & $\xi r_B+m_B =-1.04$ & \eqref{eqn:magfield}\\
            \\
            $p_{\dot M} = 2$  & $p_{\dot M} = 1 $ & \eqref{eqn:mdot}\\
            $\xi r_{\dot M}+m_{\dot M} = 4$ & $\xi r_{\dot M}+m_{\dot M} = 2.9$ & \eqref{eqn:mdot}\\
            $w = 1$  & $w =0.5$ & \eqref{eqn:wood}\\
            \\
            $p_T = 0.11  $& $p_T = 0.04 $ & \eqref{eqn:pT_ison}\\
            $\xi r_T+m_T = 0.12 $ & $\xi r_T+m_T = 0.05$ & \eqref{eqn:rmT_ison}\\
            $j=0.055$ & $j=0.02$ & \eqref{eqn:TcFx}\\
            \\
            $p_n =1.07$ & $p_n =0.64$ & \eqref{eqn:pn_ison}\\
            $\xi r_n+m_n =1.97$ & $\xi r_n+m_n =1.49$ & \eqref{eqn:rmn_ison}\\
            \noalign{\smallskip}
            \hline
         \end{tabular}
          \tablefoottext{a}{\citet{reville15b}}
          \tablefoottext{b}{\citet{matt12}}
          \tablefoottext{c}{\citet{kippenhahn}}
          \tablefoottext{d}{\citet{pizzolato}}
          \tablefoottext{e}{\citet{matt15}}
          \tablefoottext{f}{\citet{see17}}
   \end{table}
As an example, the values of all the exponents linked to the lower and upper bounds we considered in Section 3.4 (namely \(1 \leq p_B \leq 1.65\)) are given in Table \ref{tab:typical} for the single density scaling scenario (in blue in Figure \ref{scenarii}). The corresponding exponents for the two other cases can be found in Appendix E. One can notice that the configuration for which \(p_B=1.65\) ("upper bound") leads to a mass-loss rate behavior similar to \citet{holzwarth}, with \(\dot M \propto R_\star^2 F_X^{0.5}\) and \(\dot M \propto Ro^{-1}\) (\(p_{\dot M} = 1\)). However, the difference observed in Section 3.5 regarding the rotational dependency of \(T_c\) and \(n_c\) arises partially from a reasonable choice of \((\gamma,T_\odot)\). Indeed, by considering as in their study \(\gamma=1.22\) and \(T_\odot = 2.93\) MK, we obtain \(p_T = 0.074\) and \(p_n = 0.64\), which is close to their published values \(p_T=0.1\) and \(p_n=0.6\).\\
\begin{figure}[!h]
\centering
\includegraphics[scale=0.37]{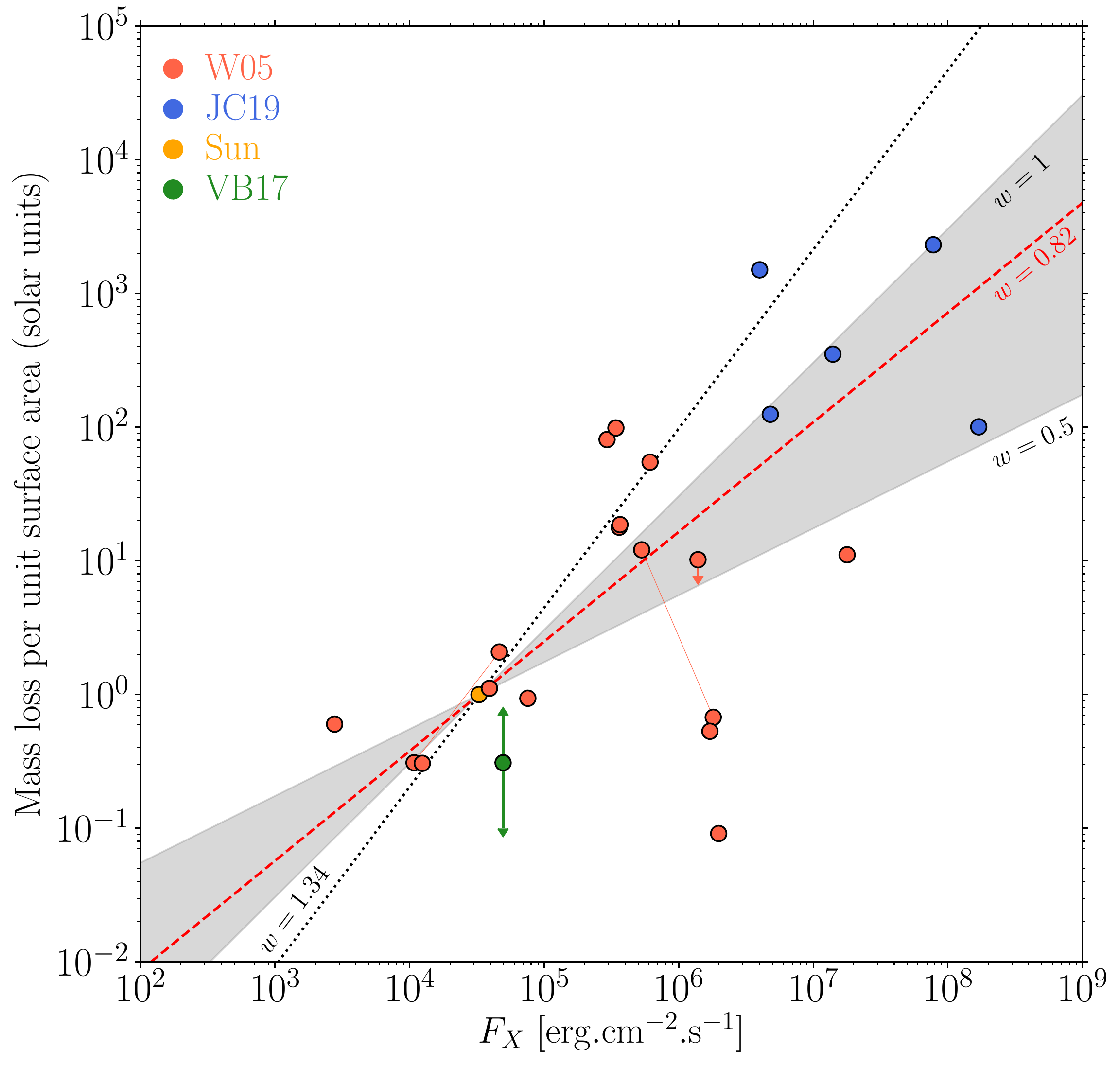}
\caption{Mass loss per unit area as a function of the X-ray flux. The red dots come from \citet{wood05} [W05] and the blue dots from \citet{jardine19} [JC19]. In green: measurements from the atmospheric evaporation of GJ 436b \citep[VB17]{vidottobourrier17}. Grey dotted line: \citet{wood05} scaling law corresponding to \(w=1.34\). Red dashed line: \citet{suzuki13} scaling law corresponding to \(w=0.82\). The light grey region corresponds to the upper and lower bounds we considered to be consistent with ZDI statistical studies.}
\label{wood}
\end{figure} 

Statistical ZDI studies, through the bounds of the \(p_B\) exponents, and prescriptions from spin-down considerations allow us to predict the admissible trends for the \(\dot M-F_X\) correlation (cf. equation \eqref{eqn:wood}). This is shown in Figure \ref{wood} as a light grey area along with the observed \(\dot M \propto R_\star^2 F_X^w\) relationship initially published by \citet{wood05}. One can notice that the predicted values of the \(w\) exponent are compatible with \citet{wood05} and \citet{jardine19} observations. Furthermore, they present a bias towards active stars with low mass-loss rates leading to weaker slopes than the \citet{wood05} prescription of \(w = 1.34\) (see the grey dotted line in Figure \ref{wood}). Nevertheless, this behavior is in agreement with \citet{suzuki13} results which predicted an exponent \(w=0.82\) (see the red dashed line in Figure \ref{wood}).

\subsubsection{Normalization factors}

If the exponents are fixed thanks to the previous calculations, the normalization of the scaling laws is still an open issue. By default, in the above analytical development, all the relevant quantities have been normalized to the solar values. However, we need to take into account additional constraints on each physical parameter in a solar configuration:

\begin{itemize}
    \item \textit{Magnetic field}. ZDI studies show a significant scatter in the dataset used to exhibit correlations between the large scale magnetic field and other stellar parameters. Such a dispersion is here taken into account in the magnetic field normalization by considering the \citet{see17} dataset, for which the average dipolar field strength at their value of the solar Rossby number lies between \(B_\odot = 0.6\text{ and }4\ \text{G}\).\\
    
    \item \textit{Mass-loss rate}. The value of \(\dot M_\odot\) is deduced from \(B_\odot\) by keeping a fixed solar wind torque, since \(\Gamma_{\text{wind},\odot} \propto B_\odot^{4m}\dot M_\odot^{1-2m}\) (cf. equation \eqref{eqn:master}). This way, for \(m=0.22\), we have \(7.9 \times 10^{-15}\leq \dot M_\odot\ [\text{M}_\odot.\text{yr}^{-1}] \leq 1.47 \times 10^{-13} \).\\
    
    \item \textit{Coronal temperature}. The normalization of the coronal temperature and density is determined by performing 1D simulations of a pressure-driven polytropic wind with \(\gamma=1.05\), using the starAML routine \citep{reville15b}. The value of \(T_\odot\) is tuned to provide an average solar wind velocity at 1 AU of 444 \(\text{km}.\text{s}^{-1}\), leading to \(T_\odot = 1.5\ \text{MK}\) \citep{reville16}.\\
    
    \item \textit{Coronal density}. The density at the base of the solar corona is then computed to be consistent with \(T_\odot\) and \(\dot M_\odot\), which results in \(2.49\times 10^7 \leq n_\odot\ [\text{cm}^{-3}] \leq 4.63\times10^8\).
\end{itemize}

Once the exponents of the scaling laws and their normalization factors are well-defined, we now can compare the different prescriptions to observations of individual systems. We will focus on ZDI studies constraining the stellar magnetic field and mass loss measurements from astrospheres' Ly\(\alpha\) absorption.
    
\subsection{A star studied through astrosphere's Ly\(\alpha\) absorption and Zeeman-Doppler Imaging: \(\epsilon\) Eridani}

We apply our formalism to \(\epsilon\) Eridani, a young active K2V dwarf which hosts an exoplanet and a debris disk. We will study this individual system by only taking the minimal information required in our formalism, in order to test the different scaling laws. In practical terms, we only rely on the stellar mass and the rotation period, which are essential to use the different prescriptions (see Table \ref{tab:EpsEri} for numerical values of those stellar parameters). An estimate of the age indicates that the star is in the Main Sequence.
\begin{table}[!h]
\centering 
      \caption{\label{tab:EpsEri} Stellar parameters of \(\epsilon\) Eridani.}
      \begin{tabular}{m{2.6cm}m{2cm}m{2.5cm}}
      \hline
      \noalign{\smallskip}
      {Star}& \multicolumn{2}{c}{$\epsilon$ Eridani}\\
      &  \multicolumn{2}{c}{Model Inputs} \\
      \noalign{\smallskip}
      \hline
    $M_\star\ (M_\odot)$ & \multicolumn{2}{c}{$0.856_{-0.008}^{+0.006\ }$\tablefootmark{a}} \\
    $P_\text{rot}\ (d)$& \multicolumn{2}{c}{11.68\tablefootmark{a}}\\
    Age (Gyr)&\multicolumn{2}{c}{0.44\tablefootmark{a}}\\
    \hline
    \noalign{\smallskip}
    &Observations&Model Outputs\\
    \noalign{\smallskip}
    \hline
    $R_{\star}\ (R_\odot)$ &$0.74\pm0.01\tablefootmark{a}$&0.87\tablefootmark{b}\\
    $L_{\star}\ (L_\odot)$&0.34\tablefootmark{c}&0.54\tablefootmark{b}\\
    $\log L_X\ (\text{erg}.\text{s}^{-1})$&28.32\tablefootmark{d}&28.24\tablefootmark{b}\\
    $Ro$&-&0.28\tablefootmark{b}\\
    \hline
    $B_\star$ (G)&6.15--19.8\tablefootmark{a,e}&3--47\tablefootmark{b}\\
    $\dot M\ (10^{-14}\ \text{M}_\odot.\text{yr}^{-1})$&30--120\tablefootmark{d}&1.9--120\tablefootmark{b}\\ 
    \noalign{\smallskip}
    \hline
  \end{tabular}
  \tablefoottext{a}{\citet{jeffers14}}
  \tablefoottext{b}{This work (Model)}
  \tablefoottext{c}{\citet{saumon}}
  \tablefoottext{d}{\citet{wood02}}
  \tablefoottext{e}{\citet{see17}}
\end{table}
  
The stellar luminosity, the stellar radius and the X-ray luminosity can be estimated through equations \eqref{eqn:LM}, \eqref{eqn:RM} and \eqref{eqn:RoLX} (see Table \ref{tab:typical} to get the associated exponents). The values obtained from those correlations, shown in Table \ref{tab:EpsEri}, are quite in agreement with their observed analogs. In practice, those relationships are crucial to estimate \(B_\star\) and \(\dot M\), since an inaccuracy in the determination of \(L_X\) would lead to an erroneous value of the mass-loss rate through the \(\dot M-F_X\) correlation.\\

\begin{figure}[!h]
\centering
\includegraphics[scale=0.375]{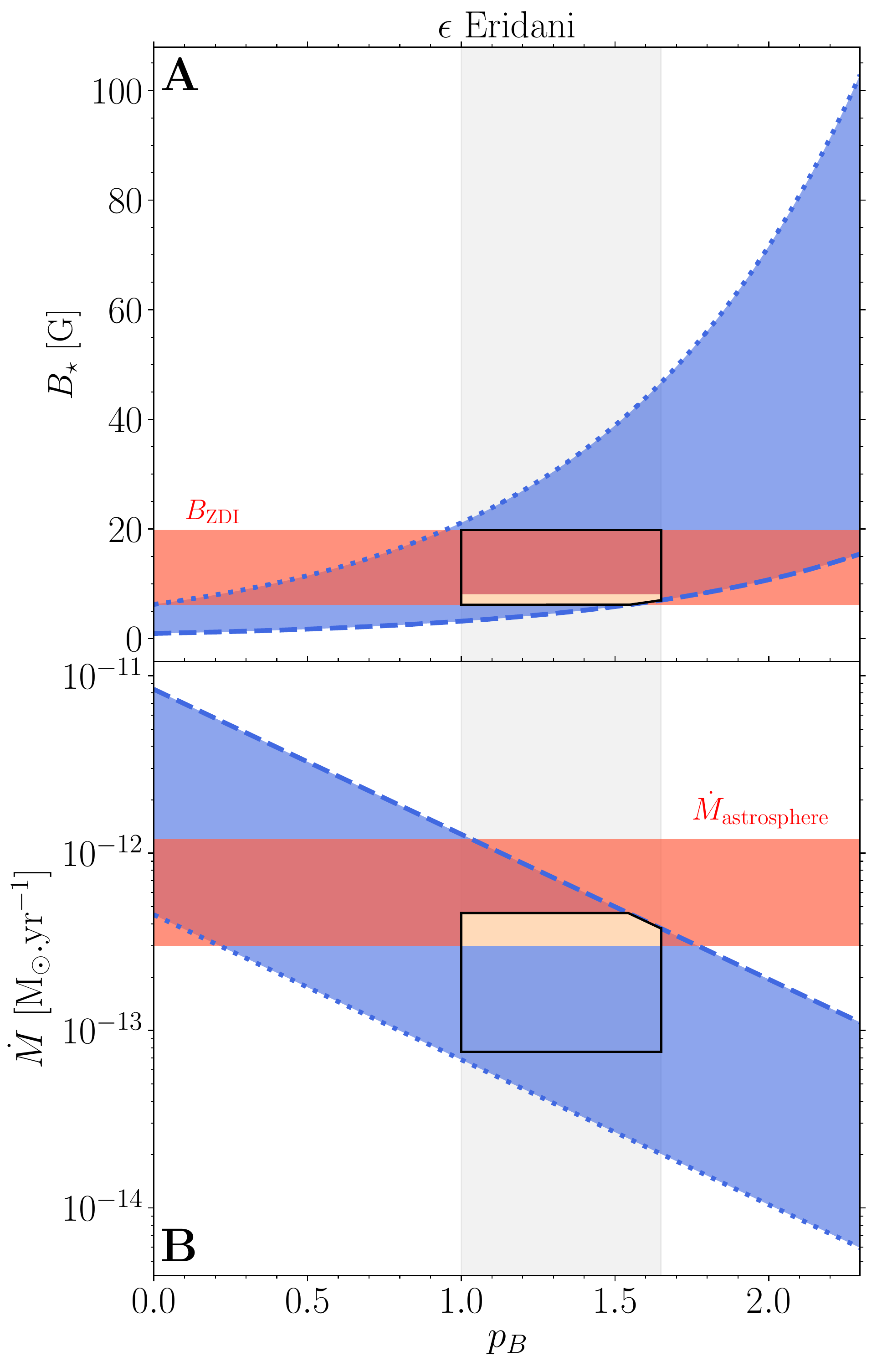}
\caption{Panel A: estimate of \(B_\star\) for \(\epsilon\) Eridani as a function of \(p_B\). Panel B: estimate of \(\dot M\) as a function of \(p_B\). In blue: predictions of our prescriptions. Dashed blue lines: normalizations corresponding to \(B_\odot = 0.6\) G. Dotted blue lines: normalizations corresponding to \(B_\odot = 4\) G. Red bands: observational constraints on the dipolar component of the magnetic field \citep{jeffers14,see17} and the mass-loss rate \citep{wood02}. The light grey band corresponds to the upper and lower bounds of \(p_B\) from large sample of ZDI studies. The black contours delimit the region of \(B_\star\) and \(\dot M\) consistent with statistical ZDI studies and individual measurements of the large-scale magnetic field. In beige: estimates consistent with all the constraints considered.}
\label{EpsEri}
\end{figure} 
All those considerations allow us to estimate the stellar magnetic field and the mass loss with our formalism. We illustrate those predictions by the blue bands in Figure \ref{EpsEri}. Note that without further assumption, our model predicts a broad range of possible \(B_\star\) (panel A) and \(\dot M\) (panel B) values. Since we are performing a systematic study of \(\epsilon\) Eridani, we will first assume that the star follows the observational trends inferred from ZDI studies. Only the exponents situated in the light grey area defined in Figure \ref{EpsEri} will be considered (\(1 \leq p_B \leq 1.65\)). A modification of the \(B_\star-Ro\) relationship then corresponds to an horizontal shift in Figure \ref{EpsEri}. A change of normalization between \(B_\odot = 0.6\) G (dashed blue line in Figure \ref{EpsEri}) and \(B_\odot = 4\) G (dotted blue line in Figure \ref{EpsEri}), linked to the scatter in the ZDI studies dataset, results in a vertical shift in Figure \ref{EpsEri}. Browsing this parameter space then gives \(3 \leq B_\star\ [\text{G}] \leq 47\) and \(1.9\times10^{-14} \leq \dot M\ [\text{M}_\odot.\text{yr}^{-1}] \leq 1.2\times 10^{-12}\).

The large-scale magnetic field of the star has been monitored for almost seven years, between January 2007 and October 2013, by \citet{jeffers14}. We only consider the dipolar component of their ZDI maps, leading to field values between 6.15 and 19.8 G \citep{see17}, thus constraining our estimates of the magnetic field  (see the red band at the top of Figure \ref{EpsEri}). Taking into account those observations, we can constrain the normalization factors and derive a narrower range for \(\dot M\): \(7.59\times10^{-14} \leq \dot M\ [\text{M}_\odot.\text{yr}^{-1}] \leq 4.59\times 10^{-13}\) (see the black contours in panel B of Figure \ref{EpsEri}).

Furthermore, \citet{wood02} measured the mass-loss rate of the star through its astrosphere's Ly\(\alpha\) absorption, leading to \(\dot M \approx 6\times 10^{-13} \ \text{M}_\odot.\text{yr}^{-1}\). Uncertainties in the determination of this value, for instance in interstellar medium properties or wind variability, introduce a systematic error of 0.3 dex in \(\log{\dot M}\), \textit{i.e.} a factor of 2 in the mass-loss rate (see \citet{wood02} for more details). Therefore, we can assume a mass loss measured through \(\epsilon\) Eridani's astrosphere between \(3\times10^{-13}\  \text{M}_\odot.\text{yr}^{-1}\) and \(1.2\times10^{-12}\ \text{M}_\odot.\text{yr}^{-1}\) (see the red band in panel B of Figure \ref{EpsEri}). This constraint further refine the acceptable parameters of our model which leads to \(B_\star\) between 6.15 and 8 G (see the beige areas in Figure \ref{EpsEri}). It is worth noticing that our \(B_\star\) and \(\dot M\) prescriptions are consistent with both statistical approaches and individual measurements (as evident in the small beige areas). Furthermore, the wind of \(\epsilon\) Eridani has been modelled in 3D with a MHD model by \citet{alvarado}. They find a mass-loss ranging from \(2.77\times 10^{-14}\) to \(10.2\times 10^{-14} M_\odot.\text{yr}^{-1}\) for January 2010. Our approach is also compatible with these values (see panel B in Figure \ref{tab:EpsEri}).\\

The range of \(\dot M\) obtained (between \(3\times 10^{-13}\) and \(4.59\times 10^{-13}\ M_\odot.\text{yr}^{-1}\)) allows us to estimate the coronal temperature \(T_c\) and the coronal density \(n_c\) for this particular star. Depending on the scenario considered (single density scaling, entropy equilibrium or single temperature scaling), since the normalization factor \(T_\odot\) has been fixed in Section 4.1.3,  the range of \(p_B\) available provides directly upper and lower bounds of \(T_c\) with our prescriptions (see Table \ref{tab:typical} and Appendix E for numerical values). The range of density is then determined by connecting the lower and upper bounds of \(\dot M\) and \(T_c\). The corresponding numerical values for the coronal properties are shown in Table \ref{tab:EpsEri_cor}.

\begin{table}[!h]
\centering 
      \caption{\label{tab:EpsEri_cor} Coronal properties of \(\epsilon\) Eridani.}
         \begin{tabu}{cccc}
            \hline
            \noalign{\smallskip}
            \text{Scenario} &$T_c\ [\text{MK}]$&$n_c\ [10^8\ \text{cm}^{-3}]$ \\
            \noalign{\smallskip}
            \hline
            \noalign{\smallskip}
            Single temperature scaling & 2.83 & 0.34--0.52\\
            Entropy equilibrium & 1.73--1.85 & 3.53--3.46\\
            Single density scaling & 1.58--1.72 & 6.98--5.62\\
            \noalign{\smallskip}
            \hline
         \end{tabu}
   \end{table}
 
One can see that the entropy equilibrium scenario yields slightly higher temperatures compared to the single density scaling hypothesis, therefore resulting in lower densities for a given mass-loss rate. In the single temperature scaling scenario, \(T_c\) is independent of the \(B_\star\) and \(\dot M\) prescriptions because
of the \(T_l-F_X\) correlation (cf. equation \eqref{eqn:TlFx}). Moreover, the star is here characterized by a low Rossby number in the unsaturated regime, leading  to a higher coronal temperature than those derived from the two other scenarii. Hence, the coronal density, in the case of a single temperature scaling, reaches the lowest values in order to keep a consistent mass loss.\\

To sum up, we have shown with \(\epsilon\) Eridani that it is possible to have an analytical prediction of the large-scale magnetic field and the mass-loss rate in agreement with all the observational constraints available, and to infer from the values obtained a range for the coronal properties of the star, according to different scenarii. Furthermore, with a systematic approach, relying on scaling laws and statistical considerations gives a quite large range for \(B_\star\) and \(\dot M\), compared to well-constrained quantities coming from individual studies. In the case of \(\epsilon\) Eridani, such a guess can deviate from the measured value by at most a factor 6 in \(B_\star\) and about one order of magnitude in \(\dot M\). If for the study of an individual star, additional measurements are required to reduce the interval of confidence, we see how powerful our approach is to guess key trends along with stellar properties in an ensemble approach. Our formalism then provides a good estimate, given our minimal set of hypothesis, of the relevant quantities from general scaling laws and statistical trends, compatible with individual studies.

\section{Conclusions and discussions}
We have provided in this paper power-law prescriptions of all the relevant parameters required to describe consistently the spin-down of solar-type stars. We confirm that the magnetic field and the mass loss are involved in a one-to-one correspondence. This is the direct consequence of assuming a generic braking torque parametrization accounting for both the distribution of stellar rotation periods in open clusters and the Skumanich law in the unsaturated rotation regime. A mass loss-X-ray flux relation coming from astropheres' Ly\(\alpha\) absorption \citep{wood05}, coupled with the knowledge of the rotational dependency of the X-ray luminosity, allowed us to link the mass-radius dependency of the aforementioned quantities to their rotational dependency. This way, we have shown that a magnetic field depending on both the Rossby number and the stellar mass may be required to remain consistent with a whole suite of observational trends. Such an approach allows us to provide upper and lower bounds for the estimates of \(B_\star\) and \(\dot M\) as follows:

\begin{itemize}
\item Lower bound:\\
\begin{equation}
B_\star\ [G] = (0.6-4)\times\left(\frac{Ro}{Ro_\odot}\right)^{-1}\left(\frac{M_\star}{M_\odot}\right)^{-1.76}
\end{equation}
\begin{equation}
\dot M\ [10^{-14}\ \text{M}_\odot.\text{yr}^{-1}] =(0.79-14.7)\times\left(\frac{Ro}{Ro_\odot}\right)^{-2}\left(\frac{M_\star}{M_\odot}\right)^4
\end{equation}
\item Upper bound:\\
\begin{equation}
B_\star\ [G] = (0.6-4)\times\left(\frac{Ro}{Ro_\odot}\right)^{-1.65}\left(\frac{M_\star}{M_\odot}\right)^{-1.04}
\end{equation}
\begin{equation}
\dot M\ [10^{-14}\ \text{M}_\odot.\text{yr}^{-1}] =(0.79-14.7)\times\left(\frac{Ro}{Ro_\odot}\right)^{-1}\left(\frac{M_\star}{M_\odot}\right)^{2.9}
\end{equation}
\end{itemize}
Furthermore, given a simple polytropic wind model and an expression of the X-ray luminosity from radiative losses, we have been able to go back to the coronal properties by assuming different scenarii linking closed loops to coronal holes. This permits us to consider in a very simplified way magnetic geometry effects occurring in stellar atmospheres. Some of these scenarii (namely scenarii 2 and 3, see \S 3.5) allow us to reconcile temperature prescriptions deduced from X-ray emission and mass-loss rate constraints, hence providing a fully consistent framework. To demonstrate the usefulness of our study, we then applied it on a real star \textit{e.g.} \(\epsilon\) Eridani. We provided estimates of the magnetic field and the mass-loss rate consistent with the different observational constraints and gave a first assessment of its coronal properties. In a saturated rotation regime, we showed that a wind torque depending linearly on the rotation rate implies a magnetic field and a mass-loss rate independent of the Rossby number. We then found a similar behavior for the coronal temperature and density, depending on the physical scenario we adopted to connect open and closed regions.\\

We managed to infer all the exponents of our scaling laws from the \(B_\star-Ro\) relation. Furthermore, in this paper we adopted an observational point of view to constrain the \(p_B\) exponent (\(B_\star \propto Ro^{-p_B}\)) by relying on the large sample of ZDI studies. One can also use theoretical dynamo scalings to determine the rotational dependency of the magnetic field \citep{augustson17a}. However, we have to bear in mind that those prescriptions are based on the magnetic energy content in the stellar interior over a wide spectral range while we considered in this work the large scale magnetic field at the stellar surface. To link the two approaches, one can fall back on a filling factor which may depend on stellar rotation \citep{see19}.

Stellar metallicity has not been directly taken into account in our study. However, it could affect significantly the coronal density and the stellar mass loss \citep{suzuki18}, thus influencing the wind braking torque. In our work, the effect of metallicity on the stellar structure in included in the Rossby dependency. Therefore, studying the influence of metallicity on the rotational evolution of solar-type stars may be a promising avenue to test different torque prescriptions.

Furthermore, coronal temperature and density are directly linked to the choice of a wind model. In this work, to connect the mass-loss rate to the coronal properties we relied on an expression for \(\dot M\) valid in the case of a non-magnetized outflow of a pressure-driven polytropic wind.
Considering a more realistic model such as a magnetized wind would introduce corrections in the mass loss expression due to the magnetocentrifugal effect \citep{preusse05,johnstone17}, involving for instance the stellar rotation rate and the Alfvén radius. It may lead to implicit relations between the different prescriptions. Given that our formalism relies on the \citet{matt15} wind braking torque, which does not take into account such an effect, a pressure-driven hydrodynamic polytropic wind may be more suitable in this context to keep a consistent model. Modifications of the wind torque and the mass-loss rate may be required to deal with very fast rotators. More complex models could also be investigated, such as a polytropic gas with a spatially varying polytropic index \citep{johnstone15a}. 

We have not considered in this paper a slow and a fast wind, which would be a way of improving even more our model.
Indeed, for a pressure-driven wind, higher temperatures lead to faster winds. However, an anti-correlation is observed between the terminal speed of the two components of the solar wind and the coronal temperature of the source region, which may be due to a difference in the altitude of the heating region between the fast and the slow component \citep{geiss, schwadron}. Therefore, more realistic wind acceleration processes, including the influence of coronal heating, have to be taken into account to deal with a fast and slow wind \citep{reville19, riley19}. 

We have shown that the \(p_L\) exponent (\(L_X \propto L_\star\ Ro^{-p_L}\)) is one of the most important parameters allowing us to constrain efficiently all the different prescriptions, especially the \(B_\star\) and \(\dot M\) mass-radius dependency as well as the expression of the coronal properties. This way, an uncertainty on this exponent \citep{pizzolato,wright} may lead to a significant scatter in our scaling laws. Therefore, our prescriptions could be significantly tightened if the interval of confidence of the \(p_L\) exponent could be reduced.

In the case of evolved stars, the decrease of the wind braking efficiency \citep{vansaders} has not been studied in this work. However, the influence of this phenomenon on the different wind parameters and its eventual inconsistency with other observational constraints may be an application of our formalism. It could be possible to introduce a \textit{re-saturation} regime at high Rossby numbers \citep{ardestani} and to look for hints of a breaking of gyrochonology in the physical quantities involved in stellar spin-down.\\


\begin{acknowledgements}
We would like to thank the anonymous referee and Sean Matt for helpful comments and suggestions regarding our work. The authors acknowledge funding from the European Union’s Horizon-2020 research and innovation programme (Grant Agreement no. 776403 ExoplANETS-A). A.S. and A.S.B. acknowledge funding by ERC WHOLESUN 810218 grant, INSU/PNST,  CNES-PLATO and CNES Solar Orbiter. A.S. acknowledges funding from the Programme National de Planétologie  (PNP). This work benefited from discussions within the international team “The Solar and Stellar Wind Connection: Heating processes and angular momentum loss”, supported by the International Space Science Institute (ISSI). We also thank Victor Réville, Manuel Güdel, Colin Johnstone, Aurélie Astoul and Kyle Augustson for useful discussions.
\end{acknowledgements}

\newpage
\begin{appendix} 
\section{Mass loss for a Parker polytropic wind}
The goal of this section is to compute the stellar mass loss by assuming a radial polytropic pressure-driven outflow, with an index \(\gamma\). We will follow \citet{lamers} in the remainder of this section.  The wind accelerates with distance and its velocity \(v\) reaches the speed of sound \(c_s\) at a critical radius \(r_c = GM_\star/2c_s^2(r_c)\) \citep{parker}. The momentum equation leads to the following integral formulation:
\begin{equation}\label{eqn:energy}
e_\gamma=\frac{v^2}{2}+\frac{c_s^2}{\gamma-1}-\frac{GM_\star}{r}=C^\text{te}.
\end{equation}
Such a constant can be estimated at the critical radius as follows:
\begin{equation}
e_\gamma=\frac{5-3\gamma}{\gamma-1}\frac{GM_\star}{4r_c}.
\end{equation}
Furthermore, from the mass conservation, we know that \(\rho v r^2\), with \(\rho\) the density of the wind, is a constant. As \(c_s^2 \propto \rho^{\gamma-1}\) for a polytrope, the speed of sound obeys to the following expression:
\begin{equation}\label{eqn:sound}
\frac{c_s^2}{c_s^2(r_c)} \propto \left(\frac{v}{c_s(r_c)}\right)^{1-\gamma}\left(\frac{r}{r_c}\right)^{2-2\gamma}.
\end{equation}
By defining \(w=v/c_s(r_c)\) and \(x=r/r_c\), equation \eqref{eqn:energy} becomes:
\begin{equation}\label{eqn:polywind}
w^{\gamma+1}-w^{\gamma-1}\left(\frac{4}{x}+\frac{5-3\gamma}{\gamma-1}\right)+\frac{2}{\gamma-1}x^{2-2\gamma}=0.
\end{equation}
\\
By definition of the critical radius, we have:
\begin{equation}
c_s^2(r_c) = \frac{GM_\star}{2r_c} = \Lambda x_0 c_s^2(R_\star),
\end{equation}
where \(x_0 = R_\star/r_c\) and \(\Lambda = v_{esc}^2/4c_s^2(R_\star)\). Therefore, thanks to equation  \eqref{eqn:sound}, assessing equation \eqref{eqn:polywind} at the base of the corona, which is assumed to be situated approximately at the stellar radius,  gives the following relationship:
\begin{equation}\label{eqn:basecorona}
\left(\Lambda x_0^{3-2\gamma}\right)^{\frac{\gamma+1}{\gamma-1}}-\left(\Lambda x_0^{3-2\gamma}\right)\left(\frac{4}{x_0}+\frac{5-3\gamma}{\gamma-1}\right)+\frac{2}{\gamma-1}x_0^{2-2\gamma}=0.
\end{equation}
If we assume that \(c_s^2(R_\star) \ll v_{esc}^2\), \textit{i.e.} \(x_0 \ll 1\) and \(\Lambda \gg 1\), we find by neglecting the first term in equation \eqref{eqn:basecorona} :
\begin{equation}
x_0 = \frac{2-4(\gamma-1)\Lambda}{(5-3\gamma)\Lambda}.
\end{equation}
At the base of the corona, we can assume that \(w\ll 1\). Therefore, equation \eqref{eqn:polywind} becomes, by neglecting the first term: 
\begin{equation}
w(R_\star)=\left[\frac{2}{\gamma-1}x_0^{3-2\gamma}\left(4+\frac{5-3\gamma}{\gamma-1}x_0\right)^{-1}\right]^{\frac{1}{\gamma-1}} = \Lambda^{\frac{1}{\gamma-1}}x_0^{\frac{3-2\gamma}{\gamma-1}}.
\end{equation}
Then the wind speed at the same distance can be expressed as:
\begin{equation}
v(R_\star) = c_s(r_c)w(R_\star) = c_s(R_\star)w(R_\star)^{\frac{\gamma+1}{2}}x_0^{\gamma-1}   .
\end{equation}
It is  now possible to estimate the mass loss as follows:
\begin{equation}
\begin{split}
\dot M &= 4\pi\rho(R_\star)R_\star^2 v(R_\star)\\
&=2\pi m_p n_c R_\star^2 c_s(R_\star) \Lambda^{\frac{\gamma+1}{2(\gamma-1)}}x_0^{\frac{5-3\gamma}{2(\gamma-1)}}.
\end{split}
\end{equation}
As \(c_s(R_\star) = \sqrt{2\gamma k_BT_c/m_p}\) for a fully ionized wind, the mass loss has the following dependencies:
\begin{equation}\label{eqn:mdotwind}
\dot M \propto \left(\frac{M_\star}{M_\odot}\right)^2 \left(\frac{n_c}{n_\odot}\right) \left(\frac{T_c}{T_\odot}\right)^{-\frac{3}{2}}\left[1-\frac{T_{\text{min},\odot}}{T_\odot}\frac{M_\star}{M_\odot}\frac{R_\odot}{R_\star}\frac{T_\odot}{T_c}\right]^{\frac{5-3\gamma}{2(\gamma-1)}},
\end{equation}
with \(T_{\text{min},\odot} = (1-1/\gamma)\ Gm_p M_\odot/2k_B R_\odot \approx 11\times 10^6\ (1-1/\gamma)\) K.

\section{X-ray luminosity from radiative losses}

The goal of this section is to compute the X-ray luminosity of the star from coronal properties. If we assume that the corona is fully ionized and optically thin, the X-ray luminosity emitted by a volume of electrons through free-free radiation can be expressed as \citep{aschwanden,see14}:
\begin{equation}
L_X = \Lambda(\bar T_l)\int_{\text{dead zone}}{n_l^2}dV \propto \Lambda(\bar T_l)\bar{n}_l^2R_\star^3\left(\frac{r_{dz}^3}{R_\star^3}-1\right),
\end{equation}
where \(\bar T_l\) and \(\bar n_l\) are respectively the mean temperature and the mean density in the dead zone, \(r_{dz}\) is the radius of the dead zone and \(\Lambda(\bar T_l)\) is the radiative loss function \citep{rosner78,aschwanden,blackman}, estimated as:
\begin{equation}
\Lambda(\bar T_l)\ [\text{erg}\ \text{cm}^3\ \text{s}^{-1}] = 10^{-17.73} \bar T_l^{-\frac{2}{3}}.
\end{equation}
This prescription is assumed to be accurate for \(\bar T_l \approx 2-10\) MK and acceptable as an average down to 0.4 MK.

To estimate the radius of the dead zone, we assume a dipolar magnetic field in the closed-field region, where the pressure gradient and the centrifugal force are not strong enough to distort the field lines. The closed loops then trap hot gas and prevent the flow of a stellar wind, allowing us to neglect the ram pressure of the gas in this area.
The edge of the dead zone is here assimilated to a limit of confinement of the plasma and therefore can be estimated with the following pressure equilibrium: 
\begin{equation}\label{eqn:defrss}
p_{th}(r_{dz}) = p_{mag}(r_{dz}),
\end{equation}
where \(p_{th}(r_{dz}),\ p_{th}(r_{dz}) \) are respectively the thermal and the magnetic pressures of the gas estimated at the edge of the dead zone. Since we consider a dipolar magnetic field, equation \eqref{eqn:defrss} becomes:
\begin{equation}
p_{th}(r_{dz}) = \frac{B_\star^2}{2\mu_0}\left(\frac{R_\star}{r_{dz}}\right)^6.
\end{equation}
If we consider that the dead zone is filled with an ideal gas of constant temperature \(\bar T_l = T_l\) and a density at the edge of the dead zone evolving similarly to the density at the base of the corona \(n_l\), \textit{i.e.} \(n_l(r_{dz}) = K n_l\), with \(K\) a constant independent of any stellar parameter, then \(p_{th}(r_{dz}) = n_l(r_{dz})k_BT_l = K n_l k_B T_l\). The radius of the dead zone then becomes:
\begin{equation}
\frac{r_{dz}}{R_\star} = \frac{r_{dz,\odot}}{R_\odot}\left(\frac{B_\star}{B_\odot}\right)^{\frac{1}{3}}\left(\frac{n_l}{n_{l,\odot}}\right)^{-\frac{1}{6}}\left(\frac{T_l}{T_{l,\odot}}\right)^{-\frac{1}{6}},
\end{equation}
where:
\begin{equation}
\frac{r_{dz,\odot}}{R_\odot} =\left(\frac{1}{K}\frac{B_\odot^2}{2\mu_0}\frac{1}{k_B n_{l,\odot} T_{l,\odot}}\right)^{\frac{1}{6}}.
\end{equation}
If we assume that \(\bar n_l\) varies the same way as the density at the base of the corona, we can express the X-ray luminosity as a function of more explicit stellar parameters:
\begin{equation}
L_X\propto \left(\frac{T_\odot}{T_l}\right)^{\frac{2}{3}}\left(\frac{n_l}{n_{l,\odot}}\right)^2\left(\frac{R_\star}{R_\odot}\right)^3\left[\left(\frac{r_{dz,\odot}}{R_\odot}\right)^3\left(\frac{B_\star}{B_\odot}\right)\left(\frac{n_{l,\odot}}{n_l}\right)^{\frac{1}{2}}\left(\frac{T_{l,\odot}}{T_l}\right)^{\frac{1}{2}}-1\right]
\end{equation}
By assuming \(r_{dz}^3/R_\star^3 \gg 1\), the X-ray luminosity becomes:
\begin{equation}\label{eqn:Lxrss_1}
L_X\propto \left(\frac{R_\star}{R_\odot}\right)^3\left(\frac{B_\star}{B_\odot}\right)\left(\frac{n_l}{n_{l,\odot}}\right)^{\frac{3}{2}}\left(\frac{T_l}{T_{l,\odot}}\right)^{-\frac{7}{6}}.
\end{equation}

\section{Conditions on the coronal temperature to generate a transsonic wind}
The goal of this section is to investigate the consequences of the positiveness of the mass-loss rate on the coronal temperature. We will assume in coronal holes a correlation between the X-ray flux and the coronal temperature \(T_c\) like in the closed loops case
\begin{equation}\label{eqn:TcFx}
T_c \propto F_X^j,
\end{equation}
where the exponent \(j\), different from \(j_l\) because of the possible difference of heating efficiency between open-field and closed-field regions, has to be determined. As for \(T_l\), we can express \(T_c\) as a function of the Rossby number and the stellar mass
\begin{equation}
T_c \propto Ro^{-p_L j}M_\star^{(\eta-2\xi)j}, 
\end{equation}
which leads to constraints similar to those obtained in \S 3.2, namely
\begin{equation}
p_T = p_L j,
\end{equation}
\begin{equation}
\xi r_T+m_T = (\eta-2\xi)j.
\end{equation}
By eliminating the \(j\) exponent, the combinated mass-radius dependency of \(T_c\) can be inferred from its Rossby dependency with
\begin{equation}
\xi r_T+m_T = \frac{\eta-2\xi}{p_L}p_T.
\end{equation}
In the case of a pressure-driven radial polytropic wind, the coronal temperature has to be greater than a threshold value to keep a positive mass-loss rate, according to equation \eqref{eqn:mdotwindfin}: 
\begin{equation}
\frac{T_c}{T_\odot}\frac{R_\star}{R_\odot}\frac{M_\odot}{M_\star}>\frac{T_{\text{min},\odot}}{T_\odot},
\end{equation}
which with equation \eqref{eqn:Tcpres} gives
\begin{equation}
\left(\frac{Ro}{Ro_\odot}\right)^{-p_T}\left(\frac{M_\star}{M_\odot}\right)^{\xi(r_T+1)+m_T-1}>\frac{T_{\text{min},\odot}}{T_\odot}.
\end{equation}
We consider here stars with \(0.5 M_\odot \leq M_\star \leq 1.4 M_\odot\) (corresponding to F, G and K spectral types) and a Rossby number lesser than \(10\ Ro_\odot\). In order to keep a well-defined mass-loss rate for all those stars, it is necessary to fulfill the following condition:
\begin{equation}
10^{-p_T}m_\text{lim}^{\xi(r_T+1)+m_T-1}>\frac{T_{\text{min},\odot}}{T_\odot},
\end{equation}
where \(m_\text{lim}\) is equal to 0.5 if \(\xi(r_T+1)+m_T-1 > 0\) and 1.4 otherwise. This will define a maximal value for the \(p_T\) exponent, considering the range of stellar masses and Rossby numbers considered:
\begin{equation}
p_T<\frac{(\xi-1)\log_{10}(m_\text{lim})-\log_{10}\left(\frac{T_{\text{min},\odot}}{T_\odot}\right)}{1-\frac{\eta-2\xi}{p_L}\log_{10}(m_\text{lim})}\equiv p_{T,\text{max}}.
\end{equation}
\begin{figure}[!h]
\centering
\includegraphics[scale=0.31]{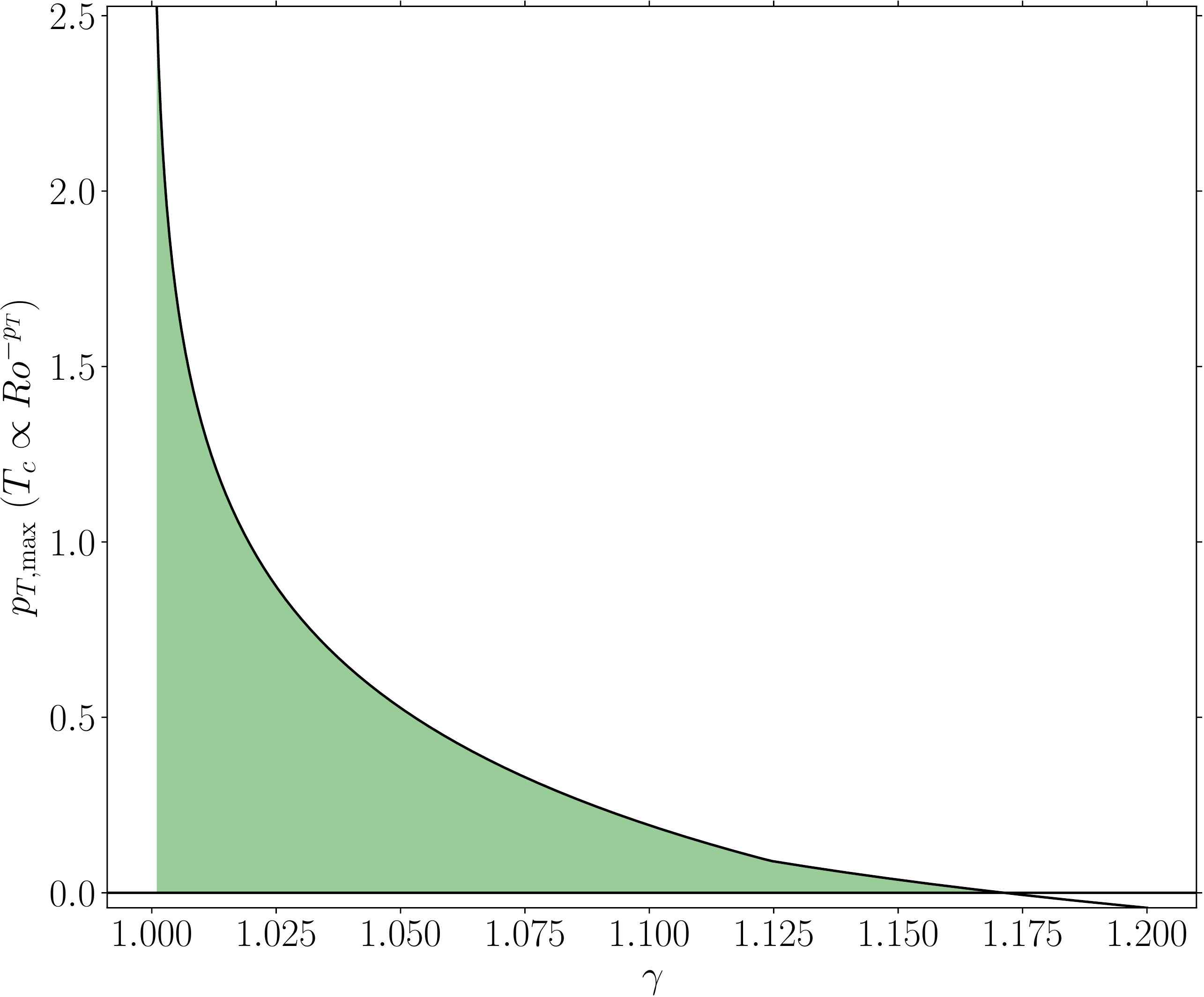}
\caption{Maximal value of \(p_T\) needed to keep a positive mass-loss rate, as a function of the adiabatic index \(\gamma\), for \(\eta = 4,\ \xi = 0.9,\ p_L = 2,\ T_\odot = 1.5\) MK. In green: consistent values of \(p_T\).}
\label{pTmax}
\end{figure}

This maximal \(p_T\) is shown in Figure \ref{pTmax}. One can notice that \(p_{T,\text{max}}\) decreases with the adiabatic index \(\gamma\) because the minimal temperature required to generate a transsonic wind increases with \(\gamma\). The \(p_{T,\text{max}}\) exponent then vanishes at a certain value of \(\gamma\), beyond which it is not possible to obtain a consistent power-law prescription of \(T_c\) without changing the value of \(T_\sun\).\\

\section{Power-law expression of the coronal density from wind model considerations}
For a pressure-driven polytropic wind, the coronal density \(n_c\) can be inferred from the mass-loss rate as
\begin{equation}
n_c \propto Ro^{-p_{\dot M}}R_\star^{r_{\dot M}}M_\star^{m_{\dot M}-2} T_c^{\frac{3}{2}}\left[1-\frac{T_{\text{min},\odot}}{T_\odot}\frac{M_\star}{M_\odot}\frac{R_\odot}{R_\star}\frac{T_\odot}{T_c}\right]^{\frac{3\gamma-5}{2(\gamma-1)}} .
\end{equation}
In order to simplify this expression, we define \(X = \frac{M_\star}{M_\odot}\frac{R_\odot}{R_\star}\frac{T_\odot}{T_c}\) and
\begin{equation}
Q(X) = \left[1-\frac{T_{\text{min},\odot}}{T_\odot}X\right]^{\frac{3\gamma-5}{2(\gamma-1)}}.
\end{equation}
To approximate this function by a power law, which can be motivated by the \(n_l\) prescription for instance, we introduce the quantity
\begin{equation}
F(\gamma,X) = \frac{\partial[\ln Q(X)]}{\partial[\ln{X}]}.
\end{equation}
The expression of the function \(Q\) then gives
\begin{equation}
F(\gamma,X) = \frac{3\gamma-5}{2(\gamma-1)}\frac{\partial\left[\ln\left(1-\frac{T_{{\text{min},\odot}}}{T_\odot}X\right)\right]}{\partial[\ln X]}.
\end{equation}

During the Main Sequence, if we assume a stellar radius approximately proportional to the mass of the star, \textit{i.e.} \(\xi \approx 1\), we can consider that \(M_\star R_\odot/M_\odot R_\star \approx 1\).  Furthermore, for small values of \(p_T\) (and small values of \(\xi r_T +m_T\) through the \(T_c-F_X\) correlation), which is required to generate a transsonic polytropic wind, we can assume that the coronal temperature presents only small variations for the stellar parameters we consider, leading to \(T_c/T_\odot \approx 1\) in \(F(\gamma,X)\). This way, \(X\approx 1\) and
\begin{equation}
F(\gamma,X) \approx \frac{5-3\gamma}{2(\gamma-1)}\frac{T_{\text{min},\odot}/T_\odot}{1-T_{\text{min},\odot}/T_\odot} \equiv F(\gamma).
\end{equation}
We can therefore express the coronal density as
\begin{equation}
n_c \propto \left(\frac{Ro}{Ro_\odot}\right)^{-p_n}\left(\frac{R_\star}{R_\odot}\right)^{r_n}\left(\frac{M_\star}{M_\odot}\right)^{m_n},
\end{equation}
with
\begin{equation}
p_n = p_{\dot M}+\left(\frac{3}{2}-F(\gamma)\right)p_T,
\end{equation}
\begin{equation}
r_n = r_{\dot M}+\frac{3}{2} r_T-F(\gamma)(1+r_T),
\end{equation}
\begin{equation}
m_n = m_{\dot M}-2+\frac{3}{2} m_T-F(\gamma)(m_T-1).
\end{equation}

\section{Exponents for other scenarii}
\begin{table}[!h]
      \centering
      \caption{\label{tab:typical_isoS} Parameters defining the \(\Gamma_\text{wind},\ B_\star,\ \dot M,\ T_c\ \&\ n_c\) prescriptions for \(p_B =1\) and \(p_B = 1.65\) in the single temperature scaling scenario and the entropy equilibrium scenario}
         \begin{tabular}{m{4cm}m{2.5cm}m{1.5cm}}
            \hline
            \noalign{\smallskip}
            \text{Lower bound} & \text{Upper bound} & \text{Equation}\\
            \noalign{\smallskip}
            \hline
            \noalign{\smallskip}
            \textit{Free parameters.}\\
            \noalign{\smallskip}
            $\gamma$ = 1.05\tablefootmark{a}& - & \eqref{eqn:mdotwindfin}\\
            $m = 0.2177\tablefootmark{d}$ & - & \eqref{eqn:torque}\\
            \noalign{\smallskip}
            \hline
            \noalign{\smallskip}
            \textit{Constrained parameters.}\\
            \noalign{\smallskip}
            $T_\odot = 1.5$ MK \tablefootmark{a} & - & \eqref{eqn:mdotwindfin}\\
            $\eta = 4\tablefootmark{b},\ \xi=0.9\tablefootmark{b}$ & - & \eqref{eqn:LM}, \eqref{eqn:RM} \\
            $p_L = 2\tablefootmark{c}$ & - &\eqref{eqn:RoLX}\\
            \noalign{\smallskip}
            $a=3.1\tablefootmark{e},\ b=0.5\tablefootmark{e}$ & - &\eqref{eqn:matt15unsat}\\
            \noalign{\smallskip}
            $p_B = 1$ & $p_B = 1.65$\tablefootmark{f}& \eqref{eqn:magfield}\\
            $\xi r_B+m_B = -1.76$ & $\xi r_B+m_B =-1.04$ & \eqref{eqn:magfield}\\
            \noalign{\smallskip}
            $p_{\dot M} = 2$  & $p_{\dot M} = 1 $ & \eqref{eqn:mdot}\\
            $\xi r_{\dot M}+m_{\dot M} = 4$ & $\xi r_{\dot M}+m_{\dot M} = 2.9$ & \eqref{eqn:mdot}\\
            $w = 1$  & $w =0.5$ & \eqref{eqn:wood}\\
            \noalign{\smallskip}
            \textit{Single temperature scaling.}&&\\
            \noalign{\smallskip}
            $p_T = 0.52  $& - & \eqref{eqn:pT_isoT}\\
            $\xi r_T+m_T = 0.57 $ & - & \eqref{eqn:rmT_isoT}\\
            $j=0.26$ & - & \eqref{eqn:TcFx}\\
            \noalign{\smallskip}
            \textit{Entropy equilibrium.}&&\\
            \noalign{\smallskip}
            $p_T = 0.17 $& $p_T = 0.11 $ & \eqref{eqn:pT_isoS}\\
            $\xi r_T+m_T = 0.19 $ & $\xi r_T+m_T = 0.13 $ & \eqref{eqn:rmT_isoS}\\
            $j= 0.09$ & $j=0.055$ & \eqref{eqn:TcFx}\\
            \noalign{\smallskip}
            $p_n = 0.55$ & $p_n =0.03$ & \eqref{eqn:pn_isoS}\\
            $\xi r_n+m_n = 1.40$ & $\xi r_n+m_n = 0.83$ & \eqref{eqn:rmn_isoS}\\
            \noalign{\smallskip}
            \hline
         \end{tabular}
          \tablefoottext{a}{\citet{reville15b}}
          \tablefoottext{b}{\citet{kippenhahn}}
          \tablefoottext{c}{\citet{pizzolato}}
          \tablefoottext{d}{\citet{matt12}}
          \tablefoottext{e}{\citet{matt15}}
          \tablefoottext{f}{\citet{see17}}
   \end{table}
\newpage
   
\cleardoublepage
\section{Observational trends used in this work and their caveats}
\begin{table}[!h]
\centering 
      \caption{Observational trends used in this work and their caveats}
         \begin{tabu}{m{2.9cm}m{4cm}m{8.7cm}}
            \hline
            \hline
            \noalign{\smallskip}
            Ingredient&References&Caveats\\
            \noalign{\smallskip}
            \hline
            \hline
            \noalign{\smallskip}
            Skumanich law&\citet{skumanich72}&\\
            &\citet{galletbouvier15}&Uncertainty in the core-envelope coupling timescale dependency. Possible break of gyrochronology for evolved stars.\\
            &\citet{vansaders}&\\
            \noalign{\smallskip}
            \hline
            \noalign{\smallskip}
            Wind braking torque&\citet{matt15}&Slows rotators with lower stellar masses spinning too fast compared to the one observed in the \textit{Kepler} field. Possible dependency on the metallicity.\\
            \noalign{\smallskip}
            \hline
            \noalign{\smallskip}
            \(L_X-Ro\) relationship&\citet{pizzolato}&\\
            &\citet{wright}&Uncertainties in the trend obtained due to observational biases.\\
            &\citet{reiners}&\\
            \noalign{\smallskip}
            \hline
            \noalign{\smallskip}
            ZDI studies & \citet{reiners12} &\\
&\citet{vidotto14}& Only large-scale unsigned magnetic flux can be measured,
missing small-scale field, observational
uncertainties.\\
&\citet{see17}&\\
            \noalign{\smallskip}
            \hline
            \noalign{\smallskip}
            Astrospheric wind
measurements&\citet{wood02,wood05}&Assumed interstellar medium
parameters, limited number of
systems, large scatter in
scaling law.\\

            \noalign{\smallskip}
            \hline
         \end{tabu}
   \end{table}
\end{appendix}

\end{document}